\documentclass[a4paper,12pt]{article}
\def\al{\alpha}
\def\be{\beta}
\def\ga{\gamma} \def\Ga{\Gamma}
\def\ep{\epsilon}
\def\lam{\lambda}

 \def\calE{{\cal E}} 
  
  \def\calL{{\cal L}}
 \def\calN{{\cal N}} \def\calO{{\cal O}}

\def\del        {  \partial  }
\def\half       {  {1\over 2}  }

\def\ie         {  {\it i.e.}      }

\def\comma          {\, ,}
\def\period         {\, .}
\def\lsim    {\lower .65ex \hbox{\ $\stackrel{<}{\sim}$\ } }
\def\gsim    {\lower .65ex \hbox{\ $\stackrel{>}{\sim}$\ } }
\def\com#1#2   { \left[#1, #2\right]} 
\def\acom#1#2  {\left\{ #1,#2\right\}}
\def\bra#1     {\langle #1 |}
\def\ket#1     {| #1 \rangle}
\def\slash#1{{\ooalign{\hfil/\hfil\crcr$#1$}}} 
\def\vecii#1#2      {  \left(\begin{array}{c}#1\\#2\end{array}\right)  }
\def\veciii#1#2#3   {  \left(\begin{array}{c}#1\\#2\\#3\end{array}
                     \right)  }
\def\veciv#1#2#3#4  {  \left(\begin{array}{c}#1\\#2\\#3\\#4
                                 \end{array}\right)  }
\def\vecfv#1#2#3#4#5 {  \left(\begin{array}{c}#1\\#2\\#3\\#4\\#5
                                 \end{array}\right)  }

\def\matrixii#1#2#3#4            {  \left(\begin{array}{cc}#1&#2\\#3&#4
                                       \end{array}\right) }
\def\matrixiii#1#2#3#4#5#6#7#8#9 {  \left(\begin{array}{ccc}#1&#2&#3\\
                                     #4&#5&#6\\#7&#8&#9\end{array}
                               \right)  }
\def\mativ#1#2#3#4               {  \left(\begin{array}{cccc}
                                       #1\\#2\\#3\\#4\end{array}\right) }
\def\matv#1#2#3#4#5              {  \left(\begin{array}{ccccc}
                                     #1\\#2\\#3\\#4\\#5\end{array}
                              \right)  }
\def\eqabegin         {  \begin{eqnarray}  }
\def\eqaend           {  \end{eqnarray}  }
\def\nn               {  \nonumber  }
\def\bracetwo#1#2     {  \left\{ \begin{array}{l} #1 \\ #2 \end{array}
                         \right.  }
\def\bracetwocases#1#2#3#4  {   \left\{ \begin{array}{ll} #1 &
                                 \qquad #2 \\
                                 #3 & \qquad #4 \end{array} \right.  }
\def\bracebegin#1     {  \left\{ \begin{array}{#1}   }
\def\braceend         {  \end{array}\right.   }
\def\parn              {  \par\noindent }

\def\parmedskip        {  \par\medskip  }

\def\parbigskipn        {  \par\bigskip\noindent  }
\def\parmedskipn        {  \par\medskip\noindent  }
\def\parsmallskipn      {  \par\smallskip\noindent  }

\def\parag#1           {\paragraph{#1} \mbox{ }\parmedskip\noindent}
\def\msection#1      {  \begin{center} \section{#1} \end{center}   }
\def\nsection#1      {  \let\boldface\bf \def\bf{} \section{#1}
                           \let\bf\boldface   }
\def\mnsection#1     {  \begin{center} \nsection{#1} \end{center}  }
\def\capsection#1    {  \let\boldface\bf \def\bf{\sc} \section{#1}
                           \let\bf\boldface   }
\def\mcapsection#1   {  \begin{center} \capsection{#1} \end{center} }

\newcommand{\nullify}[1]{}
\def\papertitlepage{\baselineskip 3.5ex \thispagestyle{empty}}
\def\Title#1{\baselineskip 1cm \vspace{1.5cm}\begin{center}
 {\Large\bf #1} \end{center} 
\vspace{0.5cm}}
\def\Authors#1{\begin{center} {\it #1} \end{center}}
\def\Abstract{\vspace{1.0cm}\begin{center} {\large\bf Abstract} 
           \end{center} \par\bigskip}
\def\Komabanumber#1#2#3{\hfill \begin{minipage}{4.2cm} UT-Komaba #1
              \parn #2 
              \parn #3 \end{minipage}}
\renewcommand{\thefootnote}{\fnsymbol{footnote}}
\renewenvironment{thebibliography}{\pagebreak[3]\par\vspace{0.6em}
\begin{flushleft}{\large \bf References}\end{flushleft}
\vspace{-1.0em}
\renewcommand{\labelenumi}{[\arabic{enumi}]\ }
\begin{enumerate}\if@twocolumn\baselineskip=0.6em\itemsep -0.2em
\else\itemsep -0.2em\fi\labelsep 0.1em}{\end{enumerate}}

\def\rdot{\dot{r}}

\def\rslash{\slash{r} }
\def\vslash{\slash{v} }

\def\thdot{\dot{\theta}}

\def\thdot{\dot{\theta}}

\newcommand{\barr}{\begin{array}}
\newcommand{\earr}{\end{array}}

\newcommand{\p}{\partial}
\def\indspace{\!\!\!\!}
\def\underscore{_}
\def\d{{\rm d}}

\def\landfz{\\[1.2ex] }

\def\brkeq{\nonumber \\[0.1ex] & }

\def\and{&}
\def\imag{i}
\def\thdot{\dot{\theta}}
\def\lf{\calL^{(4)}}
\def\lbf{\bar{\calL}^{(4)}}
\def\lt{\calL^{(2)}}
\def\depz{\delta_\ep^{(0)}}
\def\dept{\delta_\ep^{(2)}}
\def\Dept{\Delta_\ep^{(2)}}
\def\Dlamt{\Delta_\lam^{(2)}}

\def\etil{{\tilde{e}}}
\usepackage{graphicx}
\usepackage{latexsym}
\usepackage{amsmath}
\usepackage{amssymb}

\renewenvironment{thebibliography}{\pagebreak[3]\par\vspace{0.6em}
\begin{flushleft}{\large \bf References}\end{flushleft}
\vspace{-1.0em}
\renewcommand{\labelenumi}{[\arabic{enumi}]\ }
\begin{enumerate}\if@twocolumn\baselineskip=0.6em\itemsep -0.2em
\else\itemsep -0.2em\fi\labelsep 0.1em}{\end{enumerate} }

\begin{document}
\papertitlepage
\vspace*{0cm}
\Komabanumber{02-11}{hep-th/0210133} {October, 2002}
\Title{Power of Supersymmetry in  D-particle Dynamics} 
\Authors{{\sc Y.~Kazama\footnote[2]{kazama@hep3.c.u-tokyo.ac.jp} 
 and T.~Muramatsu
\footnote[3]{tetsu@hep1.c.u-tokyo.ac.jp}
\\ }
\vskip 3ex
 Institute of Physics, University of Tokyo, \\
 Komaba, Meguro-ku, Tokyo 153-8902 Japan \\
  }
\baselineskip .7cm

\numberwithin{equation}{section}
\numberwithin{figure}{section}
\numberwithin{table}{section}

\parskip=0.9ex

\Abstract

A new systematic method is developed to study to what extent the
symmetry requirements alone, above all the 
  invariance under 16 supersymmetries (SUSY),  determine the {\it
  completely off-shell} effective action $\Ga$ of a D-particle, \ie
  without imposing any restrictions on its position $r^m(\tau)$ and spin
  $\theta_\al(\tau)$.  
Our method consists of (i) writing down the proper closure relations 
 for general SUSY
transformations $\delta_\ep$ (which
necessarily involves $\Ga$ itself) together with the invariance
condition $\delta_\ep\Ga=0$  
(ii) and solving this coupled system of
functional differential equations for $\delta_\ep$ and $\Ga$
{\it simultaneously}, modulo field redefinitions, in a consistent
 derivative expansion scheme. Our analysis is facilitated by 
 a novel classification scheme introduced for 
 the terms in $\Ga$. 
At order 2 and 4, although no assumption is made on the underlying 
 theory, we reproduce the effective action 
 previously obtained at the tree and 
 the 1 loop level in Matrix theory respectively (modulo two constants), 
 together with the quantum-corrected SUSY transformations which close 
 properly. 
 This constitutes a complete 
 unambiguous proof of off-shell non-renormalization theorems.

\newpage
\baselineskip 3.5ex

\section{Introduction}  
\renewcommand{\thefootnote}{\arabic{footnote}}
One of the most remarkable discoveries with far-reaching consequences in
recent years in string/M theory is the gauge/gravity correspondence, the
prototype of which was uncovered by Maldacena \cite{maldacena}
\cite{review} in the
form of AdS/CFT duality.  Although it is certain that this phenomenon
must be deeply related to the so called $s$-$t$ duality of string
theory, \ie the modular duality between the open and the closed string
channels, it is equally certain that its validity hinges crucially on
the existence of supersymmetry: Whereas the whole tower of massive
string modes is needed at least in one of the channels for the generic
$s$-$t$ duality, the miracle of gauge/gravity correspondence in question
is that it holds without such massive modes and this can only be
possible by supersymmetric cancellations \cite{dkps}. 
More recently, the gauge/gravity relation has been extended 
 to include  the correspondence between the massive modes of the closed 
 string in a Penrose limit of $AdS_5\times S^5$ spacetime and 
 a special class of gauge-invariant composite operators in 
 $\calN=4$ super Yang-Mills theory \cite{BMN}. Although the significance
 of supersymmetry seems less apparent in this 
extension, the fact that the  Penrose limit preserves the maximal 
supersymmetry of the original  spacetime strongly suggests that 
 its role is still of considerable importance. 

In this paper, we focus on the effective dynamics of a D-particle (in
 interaction with a large number of coincident source D-particles), a
 rare example in which one can explicitly study the details of a type of
 gauge/gravity correspondence.  As is well-known, the dynamics of a
 system of D-particles is efficiently described by Matrix theory for M
 theory \cite{bfss} \cite{Susskind} \cite{Claudsonetc} and strong
 evidence has been accumulated that quantum effects in Matrix theory
 reproduce the tree-level dynamics of the 11-dimensional supergravity
 compactified along a light-like circle \cite{BeckerBecker}$\sim$\cite{HyunetalPRD}.  
Particularly impressive is the
 agreement in the case of multi-body scattering \cite{Okawa-Yoneya,%
 Okawa-Yoneya2}, which probes the non-linear structure of the
 11-dimensional supergravity.

It has been suspected that behind such remarkable successes lie the high
 degrees of symmetries of the system, in particular the maximally
 implemented supersymmetry. Indeed a number of investigations have been
 performed \cite{Pabanetal1, Pabanetal2, lowe, ss, Hyunetal, np}, which
 strongly indicated that supersymmetry, together with a few other
 symmetries, is powerful enough to fix the form of the effective action
 completely up to two constants at least at low orders in derivative
 expansions. Since the D-particle dynamics is physically non-trivial
 starting at order\footnote{The concept of order will be precisely
 defined in the next section.} 4, this evinces a rather surprising fact
 that {\it global} symmetries can be so restrictive as to dictate even
 the dynamics of the system.

However, as we have emphasized previously (see Sec.~3 of \cite{Kaz-Mura2}), 
existing analyses have several unsatisfactory features and are not 
 complete\footnote{Below we discuss only the 
most important points. 
Further remarks are provided in Sec.~2.3.}.
 The essential shortcoming is that such analyses 
made use of the eikonal approximation, or equivalently 
the tree-type on-shell condition, which sets all but $r_m, \dot{r}_m$ 
 and $\theta_\al$ to zero; higher derivatives are simply neglected. 
This is not justified for the following two related reasons: (i)\ 
Since the derivatives can be moved around by integration by parts, 
naive eikonal approximation is logically inconsistent. Derivative 
 expansion must be organized by a concept unaffected by 
 the freedom of adding total derivatives, which requires retention 
 of previously discarded terms. (ii)\ As we shall demonstrate in 
 our analysis,  fully consistent treatment  involves expressions 
 which would vanish by the quantum-corrected on-shell condition, which 
 can only be obtained by off-shell computations.  

Consequently, the only consistent and unambiguous procedure is to 
deal with the trajectory $r_m(\tau)$ and the spin degrees of freedom
 $\theta_\al(\tau)$ with {\it arbitrary} time dependence.
Based on this consideration, we have performed, in a series of 
 papers, fully off-shell 
 analysis of the D-particle system with emphasis on the role of 
 supersymmetry. After deriving the relevant 
 Ward identity \cite{Kaz-Mura1}, we computed the off-shell effective 
 action and the SUSY transformations at order 4 \cite{Kaz-Mura2},
 including all the fermionic contributions for the first time, 
proved that, given SUSY transformations,
the Ward identity uniquely fixes the effective action at this order 
\cite{Kaz-Mura3},  and finally extended this demonstration to 
 all orders in perturbation theory \cite{Kaz-Mura4}. 
These investigations were performed in the context of Matrix theory. 
Although  exceptionally powerful nature of the supersymmetry 
even at the off-shell level was revealed in these works, this was not
 sufficient to claim that SUSY determines the dynamics. One must 
 be able to show that it determines the SUSY transformations 
as well as the effective action {\it simultaneously} in a 
 self-consistent manner without any knowledge of the underlying theory
  apart from its symmetries. 
 
In this paper, we complete our program for such a demonstration. 
The basic idea is to fully utilize the  proper \lq off-shell' closure 
 relations\footnote{We elucidate what we exactly mean by 
\lq off-shell' SUSY carefully in Sec.~2.}
  that must be satisfied by the SUSY transformations $\delta_\ep$, 
 in addition to the SUSY Ward identity for the effective 
action $\Ga$. Since the expressions 
  $\delta \Ga/\delta r_m$ and $\delta \Ga/\delta \theta_\al$, 
 which would vanish on-shell, appear in the closure relations, 
 we must deal with a system of coupled non-linear functional 
 differential equations for $\delta_\ep$ and $\Ga$. This will be 
 solved in a consistent derivative expansion  with a new efficient
 classification scheme for various terms and with a careful analysis 
 of how to fix the ambiguity of adding total derivatives. 
 After a rather long analysis, with a considerable use of 
 various complicated Fierz identities, 
the following results are obtained:
\begin{enumerate}
	\item There exists a frame (\ie the choice of the definitions 
 of the fields) in which 
 the effective action and the SUSY transformations at order 2 take the 
 tree-level forms. 
	\item At order 4, the effective action is determined completely, modulo two 
 constants, which in an appropriate frame coincides with the one 
 obtained in the eikonal-type analysis\footnote{As we shall explain in 
Sec.~4.3.6, this agreement 
does not however mean that a naive eikonal analysis is justified.}
 \cite{Hyunetal} and with 
 the explicit 1-loop result in Matrix theory \cite{Kaz-Mura2}. 
	\item SUSY transformations in relevant orders are determined uniquely 
 in a chosen frame and are shown to satisfy proper closure relations. 
\end{enumerate}
It is clear that the results 1 and 2 can be interpreted as 
complete proofs of non-renormalization theorems in the context of 
Matrix theory.
 The result 3 has never been obtained before. 

Due to the large amount and to the intricate nature of the works performed in 
 this study, the exposition in this paper has become somewhat long 
 even after many omissions of the calculational details. However, 
once the basic ideas and methods described in Sec.~2 and in Sec.~4.1 are
 understood the rest of the manipulations are conceptually straightforward
 to follow. 

The organization of the rest of the article is as follows: 
We begin in Sec.~2 by describing our basic formalism. The symmetry 
 requirements are explained, including what we exactly mean by 
 off-shell supersymmetry, and basic equations are written down together 
 with our expansion scheme. Then, some salient features of our formalism 
 in comparison with previous works are clarified. 
The actual analysis begins in Sec.~3, starting at order 2. 
First the SUSY transformation
 laws are simplified by appropriate field redefinitions and a use 
 of a part of the closure relations. Then, after introducing
 a crucial concept of {\it independent basis},  the Ward identity is solved 
 and the effective action is fully determined. 
The description of our main effort, namely the analysis at order 4, 
 is given in Sec.~4. In Sec.~4.1, we introduce an efficient classification 
 scheme called \lq\lq E-type - D-type separation method",
 which at the same time 
 greatly reduces the amount of work and allows us to read off the SUSY 
 transformation laws. Using this method, we analyze the effective 
 action in Sec.~4.2 $\sim$ 4.3. Subsequently,  the SUSY transformation 
laws at this order are obtained in Sec.~4.4 and their closure 
 relations are studied in Sec.~4.5 and 4.6. 
Finally, in Section 5, we summarize our results and
indicate some directions for further study.

Two appendices are provided for some technical details. 
In Appendix A, we describe the analysis of certain special fermionic 
 transformations, called \lq null transformations', which is 
 needed to justify our scheme used in Section 4. In Appendix B, we display
   the SUSY transformation laws obtained in Sec.~4 which are
too space-filling to be presented in the main text.

\section{Basic Formalism}

\subsection{Formulation of symmetry requirements for the effective action}
A D-particle in 10 dimensions in Euclidean formulation is described by the 
9-component position vector $r^m(\tau)$ and the 16-component 
 Majorana-Weyl spinor $\theta_\al(\tau)$ representing the spin state, 
with $\tau$ the Euclidean time. 
As was already emphasized, their dependence on  $\tau$ 
 will be taken to be {\it completely arbitrary} throughout. 
The dynamics is assumed to be governed by some effective action of 
 the form 
\begin{align}
\Ga[r, \theta, g] = \int \! \d\tau \calL(r, \theta, g)\comma 
\end{align}
where $g$ is a coupling constant\footnote{
As we can easily recover its dependence from the dimensional
analysis, we will set $g = 1$ .}.
 We assign the mass dimensions $-1\comma 3\comma 1\comma
{3\over 2}$ to $\tau, g, r^m, \theta_\al$ respectively.  Thus, $\calL$
is taken to be a local expression of dimension 1. Terms composing
$\calL$ are classified according to the {\it order}, defined as the
number of time derivatives plus half the number of $\theta$'s
involved. This notion will be used to organize a consistent derivative
expansion.

We will require that $\Ga$ be invariant under (i) $SO(9)$ rotations,
 (ii) C-P-T transformations and (iii) 16 supersymmetry transformations. 
$SO(9)$ rotations act on $r^m$ and $\theta_\al$ in the usual way. 
C-P-T transformation properties are defined to conform to those valid in 
 the Matrix theory. P and T are separately violated due to the Weyl nature
 of the spinor $\theta_\al$  and we only impose invariance under C and CPT.
Under the C-transformation, $r^m \rightarrow -r^m$, while $\theta_\al$ 
 are unchanged. On the other hand, CPT-transformation does not transform
 the fields but flips the sign of the time-derivative and effects $i 
\rightarrow -i$ as it is anti-unitary. Together with the requirement of 
 hermiticity, C-P-T invariance of $\calL$ can be summarized 
 as the following simple rule \cite{Kaz-Mura3}:
\begin{itemize}
	\item In constructing $\calL$, use $i^{1+m+n} \theta^{(m)}\ga^{i_1i_2
\ldots i_k}\theta^{(n)}$ as fermionic building block, where 
 $\theta^{(m)}\equiv \del_\tau^m \theta$ and $\ga^{i_1i_2\ldots i_k}$ are
 the antisymmetrized products of $SO(9)$ $\ga$-matrices. Demand also 
 that the number of $r^m$, the number of $\ga^m$ and the \lq order' be
all even. 
\end{itemize}

Now we come to the main focus of our attention, the invariance under
 16 supersymmetries. This must be formulated and explained 
with care for several reasons. 
\begin{enumerate}
	\item As is well known, there is as yet no formulation of 16 
supersymmetries with off-shell closure:
 Commutator\footnote{Global spinor parameter $\ep_\al$ is understood to be
 included  in the transformation.} of SUSY 
transformations yields  translation only up to 
 terms which vanish upon the use of the equations of motion. As we do not
wish to  impose such on-shell conditions, we must allow for 
these additional terms in the closure relations. An obvious complication 
 is that, as they must involve $\delta \Ga/\delta r^m(\tau)$ or 
$\delta \Ga/\delta\theta_\al(\tau)$, they depend on the effective action 
itself which we wish to determine. 

\item  Sometimes, this lack of off-shell closure is rephrased as the 
 statement that for such a system 
\lq\lq supersymmetry exists only on shell". This statement
 is both true and false, depending on what one means by supersymmetry. 
If one insists that supersymmetry must act between  equal numbers of 
 bosonic and fermionic fields, then the statement is obviously correct;
imposition of the on-shell condition is indeed necessary to achieve this 
equality.  This, however, does not mean that there is no fermionic symmetry 
off the mass shell. A prime example is the super Yang-Mills theory 
 in 10 dimensions, from which the Matrix theory can be obtained by 
 dimensional reduction. For such a theory the action {\it is} invariant under 
 {\it so-called} supersymmetry transformations {\it without}
 any use of the equations of motion. A purist would carefully call it  \lq\lq a symmetry 
 which becomes the supersymmetry on-shell".  It is precisely this type of 
off-shell global fermionic symmetry that we will  impose on the effective 
 action. Having clarified its meaning, we shall hereafter simply refer to 
it as  supersymmetry, following common usage. 

\item Since we are dealing with the most general effective action 
without assuming the knowledge of the underlying theory, we must consider 
 also the most general forms for our SUSY transformation laws
 $\delta_\ep r_m$ and $\delta_\ep \theta_\al$. They are 
 to be restricted only by the generalized closure relations explained 
 above, $SO(9)$ and CPT symmetries, and dimensional considerations. 

\item We must allow arbitrary field redefinitions of the type which 
 do not change the physical S-matrix. 
\item  The actual analysis will be performed on the effective Lagrangian 
 $\calL$. Therefore we must always allow for the freedom of adding 
total derivatives. This means that a naive 
 approximation scheme, such as the often-used eikonal approximation,
 where fields with more 
than a fixed number of derivatives are set to zero is {\it not consistent}. 
On the other hand, the notion of \lq order' is stable against 
such additions. Although  we call it a \lq\lq derivative expansion", 
 what we will employ throughout is the expansion with respect to 
this quantity. 
\end{enumerate}

\subsection{Basic  equations and expansion scheme}
We are now ready to write down our basic equations which embody the
scheme explained  above.  We express the supersymmetry
transformations, their closure relations and the invariance of the
effective action under them in the following manner \footnote{Here and
hereafter, the dot signifies differentiation with respect to the Euclidean
time $\tau$ and we will use $v^m$ and $a^m$ to denote $\dot{r}^m$ and
$\ddot{r}^m$ respectively. Contractions of the spinor indices are often
suppressed, so that $(\epsilon \lambda)$ stands for $\epsilon_\beta
\lambda_\beta$, etc.}:
\begin{align}
\delta_\ep \theta_\al &= T_{\al\be}\ep_\be\comma  \label{depth}\\
\delta_\ep r_m &= \Omega_{m\be}\ep_\be\comma \label{depr}\\
\left[\delta_\ep, \delta_\lam\right] \theta_\al 
 &= -2(\ep \lam) \thdot_\al + A_{\al\be\ga\delta}
{\delta \Ga \over \delta \theta_\delta}\ep_\be \lam_\ga 
+B_{\al\be\ga n} {\delta \Ga \over \delta r_n}\ep_\be \lam_\ga\comma  
\label{clth}\\
\left[\delta_\ep, \delta_\lam\right] r_m
 &= - 2(\ep \lam) \dot{r}_m + C_{m\be\ga\delta}
{\delta \Ga \over \delta \theta_\delta}\ep_\be \lam_\ga 
+D_{m\be\ga n} {\delta \Ga \over \delta r_n}\ep_\be \lam_\ga \comma 
\label{clr}\\
\delta_\ep \Ga &= \int \! \d\tau \, \delta_\ep \calL = 0 \period \label{wardid}
\end{align}
$T, \Omega, A, B, C, D$ and $\calL$ are as yet unknown local functions
 of $\{ r_m(\tau), \theta_\al(\tau)\}$ and their derivatives.  
In what follows, $A \sim D$  will be  referred to as {\it off-shell
 coefficients}.  In the context of Matrix theory, the equation
 (\ref{wardid}) represents an invariance of the quantum effective action
 under quantum-corrected effective SUSY transformations, hence it is
 often referred to as the SUSY Ward identity or simply the Ward
 identity. In the closure relations (\ref{clth}) and 
 (\ref{clr}), we have written out the expressions
 $\delta \Ga/\delta r_m$ and $\delta \Ga/\delta \theta_\al$, which
 vanish on shell, explicitly.  We are not, however, excluding the
 possibility\footnote{Judging from the Matrix theory calculations, this
 is highly unlikely.}  that $A\sim D$ may contain additional dependence
 on functional derivatives of $\Ga$. 
Apart from the symmetry and
 dimensional requirements, the only assumption we shall make is that
 $\Ga$ starts at order 2. The prime question is to what extent the unknown
 quantities, in particular $\Ga$ and $\delta_\ep$,
 can be determined just from these
 relations, up to field redefinitions.

Let us express the basic set of equations introduced above 
in a slightly more explicit fashion. 
By using the definitions (\ref{depth}) and (\ref{depr}), the left-hand-sides 
(LHS)  of (\ref{clth}) $\sim$ (\ref{wardid}) become 
\begin{align}
\left[\delta_\ep, \delta_\lam\right] \theta_\al(\tau)
 &= \int \! \d s 
\left[\left( \Omega_{n\be}(s) {\delta T_{\al\ga}(\tau) 
\over \delta
r_n(s)} -T_{\delta\be}(s){\delta T_{\al\ga} (\tau)
 \over \delta\theta_\delta(s)} \right) +(\be \leftrightarrow \ga)
\right] \ep_\be\lam_\ga \comma 
\label{comth}
\end{align}
\begin{align}
\left[\delta_\ep, \delta_\lam\right] r_m
 &= \int  \! \d s \left[\left( -\Omega_{n\be}(s){\delta \Omega_{m\ga}(\tau)
 \over \delta r_n(s)} + T_{\al\be}(s) {\delta \Omega_{m\ga}(\tau)
 \over \delta \theta_\al(s)} \right) +(\be \leftrightarrow \ga)\right]
 \ep_\be \lam_\ga \comma 
\label{comr}\\
\delta_\ep \Ga &= \int \!\d\tau \left(\Omega_{m\be}(\tau) {\delta \Ga \over 
\delta r_m(\tau)}-T_{\al\be}(\tau) {\delta \Ga \over \delta \theta_\al(\tau)}
\right) \ep_\be \period
\label{expward}
\end{align}
This makes it clear that what we are dealing with is a set of coupled 
 non-linear functional equations, which are in general extremely hard 
 to solve. Nevertheless, by the systematic use of the  derivative 
 expansion and a novel classification scheme for the terms in $\Ga$, to be 
 described in detail later,
 one can analyze them to get concrete results at low orders. 

Now let us explain our scheme of expansion of various 
 quantities with respect to order,  needed for the analysis up to 
 order 4. 

First, the effective action, 
 the order of which must be even from CPT symmetry, is expanded as 
\begin{align}
 \Gamma & = \Gamma^{(2)} + \Gamma^{(4)}\comma 
\label{expga} 
\end{align}
where  the superscripts in  parentheses refer to their orders.
They are further expanded according to the number of $\theta$'s as 
\begin{align}
\Gamma^{(2)} & = \Gamma^{\p^2} + \Gamma^{\del\theta^2} 
+ \Gamma^{\theta^4},
\label{effexp2}\\
 \Gamma^{(4)} & = \Gamma^{\p^4} + \Gamma^{\p^3 \theta^2} +
                  \Gamma^{\p^2 \theta^4} + \Gamma^{\p \theta^6} 
+ \Gamma^{\theta^8} \label{effexp4}.
\end{align}
On the right-hand-sides (RHS), the superscript indicates the 
 schematic structure of each term in a self-explanatory manner. 

Next,  consider the   SUSY transformation laws.  A quick examination of 
 the closure relations tells us that 
$\Omega_{m\beta}$ and $T_{\alpha\beta}$ start from order 1/2 and order 1
respectively.  Also,  the order of $\Gamma$ being even, their orders
must go up by 2 units.
 Thus, we have the expansion
\begin{align}
T_{\alpha \beta} & = T_{\alpha \beta}^{(1)} + T_{\alpha \beta}^{(3)}, 
\label{expstt} \\
T_{\alpha \beta}^{(1)} & 
= T_{\alpha \beta}^{\p} + T_{\alpha \beta}^{\theta^2} , \\
T_{\alpha \beta}^{(3)} & 
= T_{\alpha \beta}^{\p^3} + T_{\alpha \beta}^{\p^2\theta^2}
+ T_{\alpha \beta}^{\p \theta^4} + T_{\alpha \beta}^{\theta^6} 
\label{expstr2}, \\[0.7ex] 
\Omega_{m \beta} &= \Omega_{m \beta}^{(1/2)} + \Omega_{m \beta}^{(5/2)}  , 
\label{expstom} \\
\Omega_{m \beta}^{(1/2)}  &= \Omega_{m \beta}^{\theta} , \\
\Omega_{m \beta}^{(5/2)}  &= \Omega_{m \beta}^{\p^2\theta} 
+ \Omega_{m \beta}^{\p \theta^3} + \Omega_{m \beta}^{\theta^5}
\label{expstth2} . 
\end{align}

Finally, consider the expansion of the off-shell coefficients. 
Again by a simple analysis of the basic equations, we find that 
 $A,B,C,D$ must start at orders
$0, {3\over 2}, {3\over 2}, 1$ respectively, and go up again by 
 2 units.  Hence their expansions become
\begin{align}
 A_{\alpha\beta\gamma\delta} & =
 A_{\alpha\beta\gamma\delta}^{(0)} +  A_{\alpha\beta\gamma\delta}^{(2)} , 
\label{expa}\\ 
 A_{\alpha\beta\gamma\delta}^{(0)} & =  A_{\alpha\beta\gamma\delta}^0  , \\
 A_{\alpha\beta\gamma\delta}^{(2)}& =  A_{\alpha\beta\gamma\delta}^{\p^2} +
A_{\alpha\beta\gamma\delta}^{\p \theta^2} 
+ A_{\alpha\beta\gamma\delta}^{\theta^4} \ , 
\end{align}
\begin{align}
B_{\alpha \beta\gamma n} & = B_{\alpha \beta\gamma n}^{(3/2)}
= B_{\alpha \beta\gamma n}^{\p \theta} +
B_{\alpha \beta\gamma n}^{\theta^3}, \\
C_{m \beta\gamma \delta} & = C_{m \beta\gamma \delta}^{(3/2)}
= C_{m \beta\gamma \delta}^{\p \theta} + 
C_{m \beta\gamma \delta}^{\theta^3}, \\
D_{m \beta \gamma n} & = D_{m \beta \gamma n}^{(1)} =
D_{m \beta \gamma n}^{\p} + D_{m \beta \gamma n}^{\theta^2}. 
\end{align}
In sections 3 and 4, we substitute these expansions into our
 basic equations, identify independent structures to produce 
more explicit set of equations and solve them. 
\subsection{Comparison with previous approaches}
Before we begin the analysis of our basic equations, it should be 
 helpful to make a comparison of our framework 
with previous  works in the literature and 
clarify what are new and/or improved in our approach. 
As we have already mentioned the essential shortcomings of 
 the eikonal-type approximation employed in existing literature in the 
 introduction, below we wish to make a little more explicit 
 comparison with the work by Paban {\it et al.}\cite{Pabanetal1}
 and the one by Hyun {\it et al.} \cite{Hyunetal}, which are most closely related 
to the present study. 

In Section 3, we shall give a complete proof of the 
 non-renormalization theorem for the effective action
at order 2, which was discussed in \cite{Pabanetal1}. 
The arguments presented in \cite{Pabanetal1} were incomplete 
 in several respects:  \ 
(i)\ It was assumed that by field redefinition the effective Lagrangian can be 
 brought to the form $f(r)v^2$ in the basis where the SUSY transformation 
 laws take the simple tree-level form without any corrections. 
As we shall see in Section 3.2, the field redefinitions which can be 
used at order 2 are actually so restricted that it is not possible to 
make {\it both} the effective action and the SUSY transformation
 laws simple at the same time. \ (ii)\ The 
$\Gamma^{\theta^4}$ term allowed 
in the effective action was neglected from the beginning. 
It requires some arguments to show that this can be eliminated. 
(iii)\ The work \cite{Pabanetal1} also discussed the determination of the 
$\Ga^{\theta^8}$ structure at order 4,  which will be dealt with
 in Sec.~4.3.4 and  4.3.5. While conditions weaker than what SUSY 
 requires were used in \cite{Pabanetal1}, we shall deal with the genuine 
 conditions dictated by SUSY. 

In Section 4, we will determine the effective action at order 4,  which 
was studied  in \cite{Hyunetal}. \ (i)\  As the authors of \cite{Hyunetal} 
 employed the  eikonal approximation, they unduly neglected 
the higher derivative terms that should be kept for consistent 
 analysis. \ (ii)\ As we shall explain in Sec.~4.2,
 these higher derivative  terms  can actually be removed by 
appropriate field redefinitions\footnote{This fact was first recognized 
by Okawa \cite{okawa9907}}. This fortunate fact does not however 
justify their treatment completely since in analyzing the Ward identity
 they again discarded higher derivative terms arbitrarily. \ 
(iii)\ Furthermore, during the course of the analysis, they 
replaced an arbitrary spinor $\epsilon_\alpha$ by a 
 special structure $ (\theta \gamma^i)_\alpha $ to simplify the 
 analysis. As a result the resultant equations provide only necessary 
 conditions. In contrast, we shall deal with the full set of constraints 
 dictated by the SUSY Ward identity. 
\ (iv)\ Finally, they only analyzed the Ward identity and did not 
 clarify the nature of the fermionic transformations. Our analysis will 
 determine the complete form of these transformations and by analyzing 
 the closure relations we shall prove that they do qualify as 
SUSY transformations. 

Having spelled out the various new features of our work in advance, 
 we now describe the essential part of the analysis. 
\section{Analysis at Order 2}
\subsection{Strategy}
We start our analysis from order 2. At this order,
 various simplifications occur and the analysis is essentially straightforward.

 The first simplification is that, by a simple counting of the 
 order,  the off-shell coefficient functions 
$B, C, D$ can be shown to vanish  at this order and we only need to keep $A$. 
Thus, the basic equations (\ref{depth}), (\ref{depr}), 
(\ref{comth}), (\ref{comr}) and  (\ref{expward}) become
\begin{align}
&\delta_\ep \theta_\al = T_{\al\be}^{(1)}\ep_\be , \\
& \delta_\ep r_m = \Omega_{m\be}^{(1/2)}\ep_\be  , 
\end{align}
\begin{align}
& \int \! \d s \left[\left( \Omega_{n\be}^{(1/2)}(s) {\delta T_{\al\ga}^{(1)}(\tau) 
\over \delta
r_n(s)} -T_{\delta\be}^{(1)}(s){\delta T_{\al\ga}^{(1)} (\tau)
 \over \delta\theta_\delta(s)} \right) +(\be \leftrightarrow \ga)
\right]  = 
- 2 \delta_{\beta\gamma} \thdot_\al 
+A^{0}_{\al\be\ga\delta} {\delta \Ga^{(2)} \over \delta \theta_\delta} \comma
\label{closureth2} \\
& \int \! \d s \left[\left( -\Omega_{n\be}^{(1/2)}(s){\delta \Omega_{m\ga}^{(1/2)}(\tau)
 \over \delta r_n(s)} + T_{\al\be}^{(1)}(s) {\delta \Omega_{m\ga}^{(1/2)}(\tau)
 \over \delta \theta_\al(s)} \right) +(\be \leftrightarrow \ga)\right]
= - 2 \delta_{\beta\gamma} \rdot_m \comma 
\label{closurer2} \\
& \hspace{3cm}
\int \! \d\tau \left(\Omega_{m\be}^{(1/2)}(\tau) {\delta \Ga^{(2)} \over 
\delta r_m(\tau)}-T_{\al\be}^{(1)}(\tau) {\delta \Ga^{(2)} \over \delta \theta_\al(\tau)}
\right) = 0  \comma 
\label{wardid2}
\end{align}
where in the last three equations 
we have removed the arbitrary spinors $\epsilon_\beta$ and
$\lambda_\gamma$. 

These equations will be solved in the following steps:
\begin{enumerate}
 \item First we write down the most general form of the SUSY transformation
laws compatible with the symmetry requirements. 
 \item Next, by utilizing the freedom of field redefinitions, 
we further simplify the form of the SUSY transformations and 
 study the restrictions from the closure relation on $r_m$. 
This will reduce $\delta_\ep$ to be of the simple tree-level form. 

 \item We then 
write down the most general expressions for the effective action,
and determine its form from the Ward identity (\ref{wardid2}).
\item Finally, we solve the closure relation (\ref{closureth2}) 
on $\theta_\alpha$ to 
determine $A_{\alpha\beta\gamma\delta}^{(0)}$.
\end{enumerate}
These steps are rather easy to perform due to several simplifying features 
 that occur at this order: Allowed structures for 
 various quantities are limited and it is not difficult to enumerate them. 
In addition, as the number of spinors is small, we need not use 
complicated Fierz rearrangement identities in solving the Ward identity. 

We now exhibit some details of the above procedures 
in the remainder of this section.

\subsection{The SUSY transformation laws and  the closure relation}

We begin by writing down the general form of the SUSY
transformation laws. As we have already described in Sec. 2,
$\Omega_{m\alpha}$ at this order is composed of  terms of
$\calO(\theta)$, while $T_{\alpha\beta}$  consists of terms of
$\calO(\p)$ and $\calO(\theta^2)$. The most general $SO(9)$ covariant 
such structures are given by 
\begin{align}
\Omega_{m\be}^{(1/2)}& =  
i (\gamma^m_{\be\gamma} \Omega_1^\theta+r_m \rslash_{\be\gamma}\Omega_2^\theta
+ r_m \delta_{\be\gamma} \Omega_3^\theta 
+ r^m \gamma^{mn}_{\beta\gamma}\Omega_4^\theta )\theta_\gamma
,  \\
T_{\al\be}^{(1)} &  = 
i\big( 
 \rslash_{\al\be}(r\cdot v)T_1^\p 
+\vslash_{\al\be}T_2^\p + \delta_{\al\be}(r\cdot v)T_3^\p 
+ \gamma^{mn}_{\al\be}r_mv_n T_4^\p  
+  T_{\alpha \beta \sigma \rho}^{\theta^2}\theta_\sigma \theta_\rho
\big) \period
\end{align}
Here $\Omega_i^\theta, T_i^\p$ $(i=1 \sim  4)$
are functions of $r(\tau)\equiv \sqrt{r^m(\tau) r^m(\tau)}$ only and
$T_{\alpha\beta\sigma\rho}^{\theta^2}$ is composed of $r^m(\tau)$ and 
$\ga$-matrices. The details of the  structure of 
$T_{\alpha\beta\sigma\rho}^{\theta^2}$ will not be needed in our analysis.

Some of the terms written above are actually forbidden  by 
 C-symmetry. The rule is that 
$\Omega_{m \beta}$ and $T_{\alpha \beta}$ must contain even and odd
number of $r^m$ (and its derivatives) respectively, since 
the tree level SUSY transformations enjoy this property and 
C-preserving quantum corrections cannot change it.
This reduces the allowed structures down to 
\begin{align}
\Omega_{m\be}^{(1/2)} & = 
 i  (\gamma^m_{\be\gamma} \Omega_1^{\theta}+r_m \rslash_{\be\gamma}\Omega_2^{\theta}
)\theta_\gamma
, \label{trans1b} \\
T_{\al\be}^{(1)} & =  
i \big(\rslash_{\al\be}(r\cdot v)
T_1^{\p} +\vslash_{\al\be}T_2^{\p} 
 +  T_{\alpha \beta \sigma \rho}^{\theta^2}\theta_\sigma \theta_\rho
\big) .
\label{trans1f}
\end{align}

Now we can further simplify these transformation laws by the use 
 of field redefinitions. The most general field redefinitions 
 that do not change the order are of the form\footnote{A possible 
 term of the form $\theta_\beta \rslash_{\beta
\alpha}$ in $\tilde \theta_\alpha$ is forbidden by C-symmetry.}
\begin{align}
\tilde r_m(r,\theta) =  r_m Z_1(r)  , \ \
\tilde \theta_\al(r,\theta) =  \theta_\al Z_2(r) \label{fredef2} \comma 
\end{align}
where $Z_i(r)$ are functions of $r(\tau)$ only. They must satisfy 
the conditions 
\begin{align}
 Z_i(r) \to 1 \ \ \  \text{as} \ \ \  r \to \infty, 
\end{align}
in order that the transformations (\ref{fredef2}) do not change the
S-matrix.  As we have two arbitrary functions $Z_i(r)$, we may \lq\lq
gauge-fix \rq\rq two functions of $r$. It is not difficult to check that
indeed we can set $\Omega^\theta_1=1$ and $\Omega^\theta_2=0$. This
choice reduces the transformation laws to
\begin{align} 
\Omega_{m\be}^{(1/2)} & = 
 i  \gamma^m_{\be\gamma} \theta_\gamma , \label{om2} \\
T_{\al\be}^{(1)} & =  
i \big(\rslash_{\al\be}(r\cdot v)
T_1^\p +\vslash_{\al\be}T_2^\p
 +  T_{\alpha \beta \sigma \rho}^{\theta^2}\theta_\sigma \theta_\rho
\big) , \label{t2}
\end{align}
where we have omitted the tilde for simplicity.

Having simplified the form of the transformation laws as much as 
 possible, let us substitute (\ref{om2}) and (\ref{t2}) into 
 the closure relation (\ref{closurer2}) on $r_m$. 
The  $\calO(\theta^0)$
and $\calO(\theta^2)$ parts of the closure relation give
\begin{align}
&  -2\,{T_2}^\p\,{v_m}\,{{\delta }_{\beta \gamma }}
   -2\,{T_1}^\p\,{r_m}\,(r \cdot v)\,
     {{\delta }_{\beta \gamma }}
= - 2\delta_{\be\ga} v_m, \label{clrth0} \\ 
& -  \left( T_{\alpha\gamma\sigma\rho}^{\theta^2} \gamma^m_{\alpha\beta} + 
T_{\alpha\beta\sigma\rho}^{\theta^2} \gamma^m_{\alpha\gamma}
\right) = 0.
\end{align}
They are easily solved and we get 
\begin{equation}
T_1^\p = 0  , \ \  T_2^\p  = 1 , \ \   
T_{\alpha\beta\sigma\rho}^{\theta^2} = 0.
\end{equation}
Thus the SUSY transformation laws  finally become 
\begin{align}
\Omega_{m\be}^{(1/2)}  =& i \gamma^m_{\beta \gamma} \theta_\gamma,  
\label{susytreer}\\
T_{\al\be}^{(1)} =& i \vslash_{\alpha \beta} 
\label{susytreeth} \period
\end{align}
What we have shown is that there exists
 a frame of fields (or a gauge) in which the SUSY transformation 
 laws at order 2 take precisely the tree-level form. 

\subsection{Determination of the 
effective action from the Ward identity}
Now we move on to the analysis of the Ward identity for the effective 
action $\Ga^{(2)}$. As we have already used up the freedom of field 
redefinitions, we must deal with the most general form of $\Ga^{(2)}$. 
By substituting
the expansion (\ref{effexp2}) of $\Ga^{(2)}$ 
 and the above SUSY transformation laws
(\ref{susytreer}), (\ref{susytreeth}) into the equation (\ref{wardid2})
and collecting terms with the same number of $\theta$'s, the Ward
identity can be split into the following three equations:
\begin{align}
& \int \! \d \tau  \left(
i 
\ga^m_{\beta\gamma}\theta_\gamma(\tau) \,  
\frac{\delta\Ga^{\p^2}}{\delta r_m(\tau)} -
i \vslash_{\alpha \beta}(\tau) \, 
\frac{\delta\Ga^{\p\theta^2}}{\delta \theta_\alpha(\tau)}
\right)  = 0 ,
\label{ward2th1} \\ 
& \int \! \d \tau  \left(
i 
\ga^m_{\beta\gamma}\theta_\gamma(\tau) \,  
\frac{\delta\Ga^{\p\theta^2}}{\delta r_m(\tau)} -
i \vslash_{\alpha \beta}(\tau) \, 
\frac{\delta\Ga^{\theta^4}}{\delta \theta_\alpha(\tau)}
\right)  = 0 , 
\label{ward2th3} \\  
& \int \! \d \tau  \, 
i 
\ga^m_{\beta\gamma}\theta_\gamma(\tau) \,  
\frac{\delta\Ga^{\theta^4}}{\delta r_m(\tau)} = 0  \, 
\label{ward2th5}.  
\end{align}
As is characteristic of any Ward identity, these equations are 
of global integrated form and due to the inherent total derivative 
 ambiguities it is non-trivial to extract the 
 information on the local quantities such as $\delta \Ga^{\del^2}/\delta
r_m(\tau)$ etc. that we wish to obtain. 

This difficulty can however be 
 overcome by the following consideration. 
First, consider the possible algebraically independent 
structures at a fixed order 
with a definite number of $\theta$'s and denote them by $\{ \etil_A\}$. 
The number $N$ of such structures is obviously finite. 
In general, there are certain number,  say $n$,  of linear combinations 
$\sum_A \tilde{g}^A_i(r) \etil_A\comma (i=1\sim n)$
 which are actually total derivatives. 
 Thus, we can choose among $\{ \etil_A\}$
 what we shall call an {\it independent basis} $\{ e_a\}_{ (a=1\sim 
N-n)}$ for which the following properties hold:
\begin{itemize}
	\item $e_a$'s are algebraically independent. 
	\item The set $\{e_a\}$ is such that $\sum g_a(r) e_a$ cannot be a 
total derivative for any choice of $g_a$'s. 
\end{itemize}
This is equivalent to the property 
\begin{align}
\int \! \d\tau \sum_a g_a(r) e_a =0 \quad \Rightarrow \quad 
 g_a(r) =0  . \label{glloc}
\end{align}
Clearly the choice of such a set $\{e_a\}$ is not unique, but once 
 we fix one independent basis and stick to it, we can unambiguously 
 obtain local equations from an integrated equation using 
 (\ref{glloc}). We must of course be very careful to check that 
a chosen set $\{e_a\}$ really satisfies these properties. For 
 structures involving more than 4 $\theta$'s, even the algebraic 
 independence can be  highly non-trivial due to the existence of 
often formidable Fierz identities. It is important to note that 
once we find an independent basis, 
(\ref{glloc}) holds for any subset of it  since it is 
 a special case with some of the $g_a$'s already set to zero. 
Hereafter, we shall say that {\it a set of terms are  \lq\lq independent" 
 whenever the property (\ref{glloc}) holds}. 
 
Using this notion of \lq\lq independence", we now solve the 
Ward identities. It is convenient to first prove $\Ga^{\theta^4} =0$. 
Of the various possible structures for $\Ga^{\theta^4}$, 
the following actually vanish by the Fierz identities:
\begin{align*}
(\theta \ga^{mn}\theta) (\theta\ga^{mn}\theta), \ \
(\theta \ga^{mnk}\theta) (\theta\ga^{mnk}\theta),  \ \
(\theta \ga^{mn}\theta) (\theta\ga^{mnk}\theta)r_k.
\end{align*}
Furthermore, by using another Fierz identity, the structure 
$(\theta \ga^{mnk}\theta)r_k (\theta\ga^{mnl}\theta)r_l$
 can be expressed in terms of the one shown just below.  
In this way,  the only possible structure for $\Ga^{\theta^4}$ is
\begin{align}
\Ga^{\theta^4} = \int \! \d\tau \, F^{\theta^4}
(\theta \ga^{an}\theta)(\theta\ga^{ak}\theta) r_n r_k, 
\label{geneff2th4} 
\end{align}
where $F^{\theta^4}$ is a function only of $r(\tau)$. 
Now substitute this into the  Ward identity (\ref{ward2th5}) and
contract it with an arbitrary spinor $\ep_\beta$. This gives 
\begin{align}
\int \! \d\tau \left(
 2(\ep \ga^m \theta)(\theta \ga^{an}\theta)(\theta \ga^{am}\theta) r_n 
 F^{\theta^4}
 + r_m (\ep \ga^m \theta) (\theta \ga^{an}\theta) (\theta \ga^{ak}\theta)
r_nr_k {\d F^{\theta^4} \over \d r} {1\over r} \right) = 0. 
\end{align}
It can be checked that the integrand does not vanish by any of 
the Fierz identities and these two terms form an independent basis.
 Thus we must set $F^{\theta^4}=0$ and hence $\Ga^{\theta^4} =0$.

With $\Ga^{\theta^4}$ eliminated,  the most general
 form of $\Ga^{(2)}$ can be written as
\begin{align}
\Gamma^{(2)} & = \Gamma^{\del^2} + \Gamma^{\del \theta^2}, \\
\Gamma^{\del^2}  & \equiv \int \! \d\tau \left(
 v^2\,{F^{\p^2}_1}+{(r \cdot v)}^2\,{F^{\p^2}_2} \right), 
\label{geneff2th0} \\
\Gamma^{\del \theta^2} & \equiv \int \! \d\tau \left(
 {r_i}\,{v_j}\,(\theta {{\gamma }^{ij}}\theta )\,{F^{\p\theta^2}_1}
   +({{\theta}}\dot \theta )\,{F^{\p\theta^2}_2} \right),
\label{geneff2th2} 
\end{align}
where $ F_i^{\p^2} $  and $ F_i^{\p\theta^2} $ ($i = 1,2$) 
are functions of $r(\tau)$ only. Here we have already discarded terms 
 forbidden by C-symmetry and those which can be eliminated by 
 integration by parts. 

As the next step, we analyze the Ward identity at 
$\calO(\epsilon \del \theta^3)$.
By substituting the expression (\ref{geneff2th2}) into (\ref{ward2th3}),
we get 
\begin{align}
\int \! \d\tau \Big( \and 
-\frac{i\,{r_i}\,{r_j}\,{v_k}\,
        (\epsilon {{\gamma }^i}\theta )\,
        (\theta {{\gamma }^{jk}}\theta )}{r}\frac{\d{F^{\p\theta^2}_1}}
      {\d r}+\frac{i\,{r_i}\,({\dot{\theta}}\theta )\,
        (\epsilon {{\gamma }^i}\theta )}{r}\frac{\d{F^{\p\theta^2}_2}}{\d r}
    \brkeq +i\,{v_i}\,(\epsilon {{\gamma }^j}\theta )\,
     (\theta {{\gamma }^{ij}}\theta )\,{F^{\p\theta^2}_1}-
    i\,{r_i}\,(\epsilon {{\gamma }^j}{\dot{\theta}})\,
     (\theta {{\gamma }^{ij}}\theta )\,{F^{\p\theta^2}_1} \Big) = 0 
\period
\label{svbth2}
\end{align}
The terms in the integrand  are already independent and hence we
must have
\begin{align}
 F^{\p\theta^2}_1 = 0, \ \  F^{\p\theta^2}_2 = c_1,
\label{solp1th3}
\end{align} 
where $c_1$ is a numerical constant.

Now we come to the analysis of the last Ward identity (\ref{ward2th1}). 
In this case, it turns out that the expression we get by the direct 
substitution of (\ref{geneff2th0}), (\ref{geneff2th2}) and 
(\ref{solp1th3}) contains dependent terms and we must perform 
 an integration by parts. In this way, (\ref{ward2th1}) can be 
brought to the form 
\begin{align}
\and \int \d \tau \Bigg(
i \Big( \frac{2}{r}\frac{\d{F^{\p^2}_1}}{\d r} -
{F^{\p^2}_2} \Big)\,
{v_i}\,(r \cdot v)\,(\epsilon {{\gamma }^i}\theta )
+ i \left(
2 {F^{\p^2}_1} - 2 c_1 \right)
\,{a_i}\,(\epsilon {{\gamma }^i}\theta )\,
-\frac{i\,v^2\,{r_i}\,(\epsilon {{\gamma }^i}\theta )}
      {r}\frac{\d{F^{\p^2}_1}}{\d r} \nn \\ 
& \hspace{1.3cm} 
- \frac{i\,{r_i}\,{(r \cdot v)}^2\,
        (\epsilon {{\gamma }^i}\theta )}{r}\frac{\d{F^{\p^2}_2}}{\d r}-
    2\,i\,{r_i}\,(r \cdot v)\,
     (\epsilon {{\gamma }^i}{\dot{\theta}})\,{F^{\p^2}_2} \Bigg) = 0\comma 
\end{align}
where the structures are now 
all independent. Thus their coefficients must separately vanish and 
we get
\begin{align}
{F^{\p^2}_1} = c_1, \ \  F^{\p^2}_2 = 0.
\end{align}

Combining all the results so far obtained, we find that 
the effective action must be of the form
\begin{align}
 \Gamma^{(2)} = \int \!\d\tau \, c_1 \left(
v^2 + (\theta \dot \theta) \right) 
\label{efftreec1}.
\end{align}
Since $c_1$ is simply a normalization constant, we will set it to $1/2$. 

What remains to be done is the examination of the closure relation 
 (\ref{closureth2}) on $\theta_\al$. By using the SUSY transformation 
laws (\ref{susytreer}), (\ref{susytreeth}) and the form of 
the effective action (\ref{efftreec1}), we easily see that the closure
 relation  fixes 
 the off-shell coefficient $A_{\alpha\beta\gamma\delta}^{0}$ to be
\begin{align}
A^{0}_{\al\be\ga\delta} = \ga^m_{\be\ga} \ga^m_{\al\delta}
 +\delta_{\be\ga}\delta_{\delta\al} -\delta_{\al\be}\delta_{\ga\delta}
 -\delta_{\al\ga} \delta_{\be\delta} .
\label{A0}
\end{align}
%
\subsection{Summary of the results at order 2}

Let us pause to summarize the results found at order 2. 
What we have shown is that at this order the symmetry requirements 
are powerful enough to fix the effective action and the SUSY transformation
 laws completely in such a manner that the proper 
closure relations are fulfilled. In an appropriate frame, 
they take the simple tree-level form 
\begin{align}
 \Gamma^{(2)} &= \int \!\d\tau \, \half \left( v^2 + (\theta
\dot \theta) \right), 
\label{efforder2}\\ 
\Omega_{m\be}^{(1/2)} &= i \gamma^m_{\beta
\gamma} \theta_\gamma, 
\label{susytreersum}
\\ 
T_{\al\be}^{(1)} &= i \vslash_{\alpha \beta} \label{susytreethsum}.
\end{align}
The analysis was completely non-perturbative and it can be interpreted as
an unambiguous proof of a non-renormalization theorem in the 
 context of Matrix theory for M theory. 
\section{Analysis at Order 4}

The analysis at order 4 is considerably more involved due to a vast
 number of possible structures and to the need of often formidable Fierz
 identities. We shall overcome the essential part of 
this difficulty by devising  a novel classification scheme for various 
terms that occur in the effective action. The basic idea is to 
 separate, within a given order, the type of terms which occur 
in the naive eikonal approximation and the rest containing more derivatives. 
Combined with judicious field redefinitions and the use of the 
notion of \lq\lq independent basis" already described, we can reduce the 
amount of analysis considerably to be able to solve our basic 
 equations (\ref{depth}) $\sim$ (\ref{wardid}) completely. 
This method, to be described in detail below, has a further advantage that 
we can obtain the SUSY transformation laws rather easily. 

\subsection{Scheme of the analysis}
Since the actual process of solving the basic equations is somewhat 
complicated, we spell out, in this subsection, 
 the essence of our scheme of analysis. 
\parbigskipn
{\bf E-type - D-type separation and simplification of the effective 
action}
\parmedskipn
First, we classify each term that may occur in the effective action 
 into {\it E-type} and {\it D-type},  defined as follows:
\begin{itemize}
	\item E-type:\quad An expression involving $r_m, v_m$ and
		      $\theta_\al$ only will be called
of eikonal- or E-type. 
	\item D-type:\quad An expression containing higher derivatives, such
		      as $a_m, \thdot_\al$ etc.,
 will be called of derivative- or D-type.
\end{itemize}

Using this terminology,  the effective Lagrangian 
$\lf$ at order 4 can be written as a sum of 
 an E-type part $\lbf$ and the rest forming a D-type part in the
following way:
\begin{align}
\lf \simeq \lbf + a_m X_m -\Psi_\al\thdot_\al  \period
\label{eff2ax}
\end{align}
Here, the symbol $\simeq$ signifies equality up to a total derivative, 
and $X_m$ and $\Psi_\al$ are arbitrary expressions of order 2 and $5/2$ 
respectively. It should be clear that the D-type part can always be 
 brought to the form   above by adding appropriate total derivatives. 
Obviously this E-D separation is not unique: 
An E-type term in $\lbf$ containing
 $v_m$ can be rewritten, by \lq\lq integration by parts", 
into sum of E-type and D-type terms. As we shall explicitly demonstrate 
in Sec.~4.2, this ambiguity can be completely eliminated by first 
fixing a complete basis for $\lf$ and then choosing among them an 
independent basis for $\lbf$, $X_m$ and $\Psi_\al$. 
Here we suppose that such a basis has been chosen. 

Now we make use of the observation by Okawa \cite{okawa9907}
 that the D-type terms in (\ref{eff2ax}) can be removed by the following field
redefinitions applied to the Lagrangian $\lt$ at order 2:
\begin{align}
r_m &\rightarrow   r_m + X_m \comma \\ 
\theta_\al &\rightarrow \theta_\al + \Psi_\al \period 
\end{align}
Indeed, one can easily check that, up to total derivatives, the extra
terms produced from $\lt$ through these field redefinitions 
cancel the D-type terms of $\lf$. 
Thus,  {\it $\lf$ can be brought to a
form consisting only of E-type terms. }
We will schematically write $\lbf$ as $\lbf =\sum_i f_i(r) e_i$, where
 $\{e_i\}$ is a basis of E-type terms and $f_i(r)$ are the 
 coefficient functions. 
\parbigskipn
{\bf Procedure for the analysis of the Ward Identity}

Next we will examine the Ward identity at order 4. As we have already made use 
 of field redefinitions, we must deal with the most general form of the 
SUSY transformations. 
Denoting such transformation at order 0 and 2 
by $\depz$ and $\dept$ respectively, the Ward identity 
is expressed as
\begin{align}
0 \simeq \depz \lbf +\dept \lt .
\label{ward24}
\end{align}
Consider the first term. Since $\lbf$ has terms containing $v_m$, 
the action of $\depz $  on $v_m$ produces terms
 with one $\thdot$, which are of D-type. Hence, $\depz \lbf$ is 
of the structure $\bar{E}[f]+\bar{D}[f]$, where $\bar{E}[f]$ and 
$\bar{D}[f]$ denote schematically
 the  E-type  and the D-type terms respectively, which depend on 
the coefficient functions $f_i$.  If the terms in $\bar{E}[f]$
 are not all independent, we rewrite the non-independent terms as much as
 possible into D-type terms using integration by parts. After this 
 manipulation, we get 
\begin{align}
 \depz \lbf \simeq  E[f]+D[f] \comma 
\label{del0L4}
\end{align}
where $E[f]$ here contains independent structures only. 
Furthermore, it is important to recognize that the terms composing
 $D[f]$ are actually of special type. As they are produced either from 
the variation of $v_m$, as already explained, or from partial integration
 of E-type terms, they can only contain one
$\dot\theta_\alpha$ or one $a^m$. Thus, $D[f]$
must be of the form 
\begin{align}
 D[f] = a^m \calE_m[f] + \calE_\alpha[f] \dot \theta^\alpha  ,
\label{Dofdel0L4}
\end{align}
where $\calE_m[f]$ and $\calE_\alpha[f]$ are schematic expressions for
bosonic and fermionic E-type terms respectively,  which are functions of 
$f_i$. 

Now consider the second term of the Ward identity (\ref{ward24}),  namely
 $\dept \lt$. 
Due to the form of $\lt$, it can be brought to the form
\begin{align}
\dept \lt \simeq -a_m \dept r_m[h]  + \dept \theta_\al \thdot_\al[h]  ,
\label{del2L2}
\end{align}
where $h=\{h_k\}$ collectively denotes the coefficient functions for the 
structures that can appear in the SUSY transformation laws. 
Evidently,  $\dept \lt$ consists only of D-type terms, which we denote
 by $\{\bar d_i\}$. Recalling that $\dept r_m$ and $\dept \theta_\al$
 are still arbitrary, this set contains the special type of 
 D-type terms composing $D[f]$ above. 

Now an important question is whether the set $\{\bar d_i\}$ forms 
 an independent basis for D-type terms. The answer would be \lq\lq no" if 
there exists some SUSY transformation 
$\Delta^{(2)}_\epsilon$,  referred to as  ``null''
transformation, for which the RHS of (\ref{del2L2}) becomes a 
total derivative, \ie 
\begin{align}
-a_m \Delta^{(2)}_\epsilon r_m + \Delta^{(2)}_\epsilon 
\theta_\al \thdot_\al =  \frac{\d G}{\d \tau} ,
\label{nullequation}
\end{align}
for some  $G$. A detailed investigation of this equation, summarized in 
 Appendix A, shows that although such null transformations exist, 
 they cannot satisfy the proper SUSY closure relations and therefore should 
 be excluded. This proves that the set  $\{\bar d_i\}$ forms an  independent
basis for D-type terms.

Combining the results (\ref{ward24}), (\ref{del0L4}), (\ref{Dofdel0L4}) and
(\ref{del2L2}), the Ward identity can be written as 
\begin{align}
E[f]+a^m  \calE_m[f]+ \calE_\alpha[f] \dot \theta^\alpha  
-a_m \dept r_m[h] + \dept \theta_\al \thdot_\al[h] 
 \simeq 0\comma 
\end{align}
and, as it is expressed in terms of independent basis, it leads to 
 the following set of local
 equations:
\begin{align}
E[f] &= 0  \label{e0}, \\ 
\calE_m[f]  - \dept r_m[h] &= 0 ,  \label{dd1} \\ 
\calE_\alpha[f] + \dept \theta_\al[h] &=0  . \label{dd2}
\end{align}
The first equation imposes relations among $f_i$'s and, as we shall see, 
determines the form of the effective action. The second and the third 
equations, on the other hand, will enable us to express $h_k$ in terms 
 of $f_i$, thereby determining the form of the SUSY transformations
 directly. This feature is extremely useful since it spares us 
of enumerating all possible SUSY transformations, a task of 
 considerable  complexity. 
\parbigskipn
{\bf Analysis of the Closure relations}

Finally, using the effective action and the transformation laws 
 thus obtained, we examine the closure equations to 
 prove that these transformation laws truly qualify as those of supersymmetry.
 This type of analysis has never been performed before and 
 it at the same time determines the form of the off-shell coefficient 
 functions $A \sim D$ completely.

In what follows, we will describe in some detail 
 how the procedures sketched above are actually executed .

\subsection{General form of the effective action}
First, we must write down the most general form of the  effective 
 action $\Ga^{(4)}$. Performing appropriate integration by parts to 
the expressions already obtained in our previous 
 work \cite{Kaz-Mura3} to bring it to the \lq\lq standard form" 
(\ref{eff2ax}), we can write it as 
\begin{align}
\Ga^{(4)}  =  \Gamma^{\p^4} + & \Gamma^{\p^3 \theta^2} + 
\Gamma^{\p^3 \theta^2}+\Gamma^{\p \theta^6}+\Gamma^{\theta^8} , \ \\
\Gamma^{\p^4}  =  \int \! \d\tau  \Big( \and 
{{{f^{{{\partial}^4}}}}_{\indspace 1}}\,v^4
+
    {{{F^{{{\partial}^4}}}}_{\indspace 2}}\,
     {({r \cdot a})}^2+{{{F^{{{\partial}^4}}}}_
       {\indspace 3}}\,({r \cdot v})\,({v \cdot a}) \brkeq  
  +{{{F^{{{\partial}^4}}}}_{\indspace 4}}\,({r \cdot a})\,
     v^2
+
{{{F^{{{\partial}^4}}}}_{\indspace 5}}\,
     {({r \cdot v})}^2\,({r \cdot a})    
+{{{F^{{{\partial}^4}}}}_{\indspace 6}}\,a^2  \Big) , \ \\
\Gamma^{\p^3 \theta^2} =  \int \! \d\tau  \, \Big(
  \and 
    {{{f^{{{\partial}^3}{{\theta}^2}}}}_{\indspace 1}}\,v^2\,{v_{{i}}} 
     {r_{{j}}}\,(\theta {{\gamma }^{{i}{j}}}\theta )
+
    {{{F^{{{\partial}^3}{{\theta}^2}}}}_{\indspace 2}}\,
     {({r \cdot v})}^2\,({\dot{\theta}}\theta )+
    {{{F^{{{\partial}^3}{{\theta}^2}}}}_{\indspace 3}}\,
     ({r \cdot a})\,({\dot{\theta}}\theta )
   +{{{F^{{{\partial}^3}{{\theta}^2}}}}_{\indspace 4}}\,
     ({\dot{\theta}}\theta )\,v^2 \brkeq 
+
    {{{F^{{{\partial}^3}{{\theta}^2}}}}_{\indspace 5}}\,
     {r_{{j}}}\,({\dot{\theta}}{{\gamma }^{{i}{j}}}
      {\dot{\theta}})\,{v_{{i}}}  
+
    {{{F^{{{\partial}^3}{{\theta}^2}}}}_{\indspace 6}}\,
     ({r \cdot v})\,{r_{{j}}}\,
     ({\dot{\theta}}{{\gamma }^{{i}{j}}}\theta )\,
     {v_{{i}}} 
   +{{{F^{{{\partial}^3}{{\theta}^2}}}}_{\indspace 7}}\,
     ({r \cdot a})\,{r_{{j}}}\,
     (\theta {{\gamma }^{{i}{j}}}\theta )\,{v_{{i}}}
\brkeq 
+ 
{{{F^{{{\partial}^3}{{\theta}^2}}}}_{\indspace 8}}\,
     ({\ddot{\theta}}{\dot{\theta}})
+{{{F^{{{\partial}^3}{{\theta}^2}}}}_
       {\indspace 9}}\,{r_{{j}}}\,
     ({\dot{\theta}}{{\gamma }^{{i}{j}}}\theta )\,
     {a_{{i}}} 
   +{{{F^{{{\partial}^3}{{\theta}^2}}}}_{\indspace 10}}\,
     ({r \cdot v})\,{r_{{j}}}\,
     (\theta {{\gamma }^{{i}{j}}}\theta )\,{a_{{i}}} \brkeq 
+
    {{{F^{{{\partial}^3}{{\theta}^2}}}}_{\indspace 11}}\,
     (\theta {{\gamma }^{{i}{j}}}\theta )\,{v_{{i}}}\,
     {a_{{j}}} \Big) ,  
\end{align}
\begin{align}
\Gamma^{\p^2 \theta^4} =  \int \! \d\tau  \, \Big( \and 
f^{{\p}^2{\theta}^4}\underscore1\,v^2\,{r_i}\,
     {r_j}\,(\theta {{\gamma }^{ik}}\theta )\,
     (\theta {{\gamma }^{jk}}\theta )+
    f^{{\p}^2{\theta}^4}\underscore2\,{v_i}\,{v_j}\,
     (\theta {{\gamma }^{ik}}\theta )\,
     (\theta {{\gamma }^{jk}}\theta ) \brkeq 
+f^{{\p}^2{\theta}^4}\underscore3\,{r_i}\,{r_j}\,
     {v_k}\,{v_l}\,(\theta {{\gamma }^{ik}}\theta )\,
     (\theta {{\gamma }^{jl}}\theta ) 
+ 
  {F}^{{\p}^2{\theta}^4}\underscore{4}\,{r_i}\,{r_k}\,
   (r \cdot a)\,(\theta {{\gamma }^{ij}}\theta )\,
   (\theta {{\gamma }^{kj}}\theta )  \brkeq 
+ 
  {F}^{{\p}^2{\theta}^4}\underscore{5}\,{r_k}\,{a_i}\,
   (\theta {{\gamma }^{ij}}\theta )\,
   (\theta {{\gamma }^{kj}}\theta ) 
+ 
  {F}^{{\p}^2{\theta}^4}\underscore{6}\,{r_j}\,{v_k}\,
   (\theta {{\gamma }^{ik}}\theta )\,
   ({\dot{\theta}}{{\gamma }^{ij}}\theta )  \brkeq + 
  {F}^{{\p}^2{\theta}^4}\underscore{7}\,
   {({\dot{\theta}}\theta )}^2 + 
  {F}^{{\p}^2{\theta}^4}\underscore{8}\,{r_i}\,{v_j}\,
   ({\dot{\theta}}\theta )\,(\theta {{\gamma }^{ij}}\theta ) 
    + {F}^{{\p}^2{\theta}^4}\underscore{9}\,{r_j}\,{v_k}\,
   (\theta {{\gamma }^{ij}}\theta )\,
   ({\dot{\theta}}{{\gamma }^{ik}}\theta )  \brkeq + 
  {F}^{{\p}^2{\theta}^4}\underscore{10}\,{r_j}\,{r_k}\,
   (r \cdot v)\,(\theta {{\gamma }^{ij}}\theta )\,
   ({\dot{\theta}}{{\gamma }^{ik}}\theta ) + 
  {F}^{{\p}^2{\theta}^4}\underscore{11}\,
   {({\dot{\theta}}{{\gamma }^{i}}\theta )}^2  \brkeq + 
  {F}^{{\p}^2{\theta}^4}\underscore{12}\,{r_i}\,{r_j}\,
   ({\dot{\theta}}{{\gamma }^{i}}\theta )\,
   ({\dot{\theta}}{{\gamma }^{j}}\theta )  + 
  {F}^{{\p}^2{\theta}^4}\underscore{13}\,{r_j}\,{r_k}\,
   (\theta {{\gamma }^{ij}}\theta )\,
   ({\dot{\theta}}{{\gamma }^{ik}}{\dot{\theta}}) \Big) , \\
\Gamma^{\p \theta^6} =  \int \! \d\tau  \, \Big(
  \and f^{\p{\theta}^6}\underscore1\,{r_i}\,{v_j}\,
     (\theta {{\gamma }^{il}}\theta )\,
     (\theta {{\gamma }^{jk}}\theta )\,
     (\theta {{\gamma }^{kl}}\theta )
+f^{\p{\theta}^6}\underscore2\,{r_i}\,{r_j}\,{r_k}\,
     {v_l}\,(\theta {{\gamma }^{im}}\theta )\,
     (\theta {{\gamma }^{jm}}\theta )\,
     (\theta {{\gamma }^{kl}}\theta ) \brkeq 
{F}^{\p{\theta}^6}\underscore3\,{r_i}\,{r_j}\,
     ({\dot{\theta}}\theta )\,
     (\theta {{\gamma }^{ik}}\theta )\,
     (\theta {{\gamma }^{jk}}\theta )
+ 
    {F}^{\p{\theta}^6}\underscore4\,{r_i}\,{r_j}\,
     (\theta {{\gamma }^{ik}}\theta )\,
     (\theta {{\gamma }^{kl}}\theta )\,
     ({\dot{\theta}}{{\gamma }^{jl}}\theta )  \Big) , \\
 \Gamma^{\theta^8}  =  \int \! \d\tau \, \Big( &
f^{{\theta}^8}\underscore1\,
     (\theta {{\gamma }^{ij}}\theta )\,
     (\theta {{\gamma }^{ik}}\theta )\,
     (\theta {{\gamma }^{jl}}\theta )\,
     (\theta {{\gamma }^{kl}}\theta ) \nn \\  & \hspace{2cm} 
   +f^{{\theta}^8}\underscore2\,{r_i}\,{r_j}\,
     (\theta {{\gamma }^{ik}}\theta )\,
     (\theta {{\gamma }^{jm}}\theta )\,
     (\theta {{\gamma }^{kl}}\theta )\,
     (\theta {{\gamma }^{lm}}\theta ) \nn \\  & \hspace{3cm} 
   +f^{{\theta}^8}\underscore3\,{r_i}\,{r_j}\,{r_k}\,
     {r_l}\,(\theta {{\gamma }^{im}}\theta )\,
     (\theta {{\gamma }^{jm}}\theta )\,
     (\theta {{\gamma }^{kn}}\theta )\,
     (\theta {{\gamma }^{ln}}\theta )
\Big) ,
\end{align}
where the coefficients $f_i$'s (for E-type terms) and $F_i$'s (for
D-type terms) are functions of $r(\tau)$ only.

 As already explained,  we can
remove all the D-type terms contained in this expression by appropriate
field redefinitions. Consider for example the case of $\Gamma^{\p^4}$,
 where all but the first term are of D-type. It is easy to verify that
they can be removed by 
the following field redefinitions applied to $\calL^{(2)}$:
\begin{align}
r^m  \ \ \longrightarrow \ \ 
\tilde{r}^m  = & r^m +  
{{{F^{{{\partial}^4}}}}_{\indspace 2}}(r)\,
     {({r \cdot a})} r^m  
+{{{F^{{{\partial}^4}}}}_
       {\indspace 3}}(r)\,({r \cdot v})\,v^m  \brkeq
   +{{{F^{{{\partial}^4}}}}_{\indspace 4}}(r)\,
     v^2 r^m +
{{{F^{{{\partial}^4}}}}_{\indspace 5}}(r)\,
     {({r \cdot v})}^2\,r^m 
   +{{{F^{{{\partial}^4}}}}_{\indspace 6}}(r)\,a^m  .
\end{align}
In a similar manner, all the D-type terms in $\Gamma^{\p^3 \theta^2}$,
$\Gamma^{\p^2\theta^4}$, $\Gamma^{\p \theta^6}$ can be eliminated.
 The simplified
effective action, consisting only of E-type terms,  contains terms 
 of the form 
\begin{align}
 \Gamma^{\p^4} & = \int \! \d\tau \, f^{{\p}^4}\underscore1 v^4, 
\label{eff4p4th0} \\
  \Gamma^{\p^3 \theta^2} & =  \int \! \d\tau \,
 f^{{\p}^3{\theta}^2}\underscore1\,v^2\,{r_j}\,{v_i}\,
  (\theta {{\gamma }^{ij}}\theta ), \label{eff4p3th2} \\
 \Gamma^{\p^2 \theta^4} & =  \int \! \d\tau \, \left(
f^{{\p}^2{\theta}^4}\underscore1\,v^2\,{r_i}\,
     {r_j}\,(\theta {{\gamma }^{ik}}\theta )\,
     (\theta {{\gamma }^{jk}}\theta )+
    f^{{\p}^2{\theta}^4}\underscore2\,{v_i}\,{v_j}\,
     (\theta {{\gamma }^{ik}}\theta )\,
     (\theta {{\gamma }^{jk}}\theta ) \right. \nn \\
& \left. \hspace{2cm}  +f^{{\p}^2{\theta}^4}\underscore3\,{r_i}\,{r_j}\,
     {v_k}\,{v_l}\,(\theta {{\gamma }^{ik}}\theta )\,
     (\theta {{\gamma }^{jl}}\theta ) 
\right), \label{eff4p2th4}  
\end{align}
\begin{align}
\Gamma^{\p \theta^6} & =  \int \! \d\tau \, \left(
 f^{\p{\theta}^6}\underscore1\,{r_i}\,{v_j}\,
     (\theta {{\gamma }^{il}}\theta )\,
     (\theta {{\gamma }^{jk}}\theta )\,
     (\theta {{\gamma }^{kl}}\theta ) 
\right. \nn \\ 
& \hspace{2cm} 
\left.    +f^{\p{\theta}^6}\underscore2\,{r_i}\,{r_j}\,{r_k}\,
     {v_l}\,(\theta {{\gamma }^{im}}\theta )\,
     (\theta {{\gamma }^{jm}}\theta )\,
     (\theta {{\gamma }^{kl}}\theta ) \right), \label{eff4p1th6} \\
\Gamma^{\theta^8} & =  \int \! \d\tau \, \Big(
f^{{\theta}^8}\underscore1\,
     (\theta {{\gamma }^{ij}}\theta )\,
     (\theta {{\gamma }^{ik}}\theta )\,
     (\theta {{\gamma }^{jl}}\theta )\,
     (\theta {{\gamma }^{kl}}\theta ) \nn \\  & \hspace{2cm} 
   +f^{{\theta}^8}\underscore2\,{r_i}\,{r_j}\,
     (\theta {{\gamma }^{ik}}\theta )\,
     (\theta {{\gamma }^{jm}}\theta )\,
     (\theta {{\gamma }^{kl}}\theta )\,
     (\theta {{\gamma }^{lm}}\theta ) \nn \\  & \hspace{3cm} 
   +f^{{\theta}^8}\underscore3\,{r_i}\,{r_j}\,{r_k}\,
     {r_l}\,(\theta {{\gamma }^{im}}\theta )\,
     (\theta {{\gamma }^{jm}}\theta )\,
     (\theta {{\gamma }^{kn}}\theta )\,
     (\theta {{\gamma }^{ln}}\theta )
\Big) . \label{eff4p0th8}
\end{align}

\subsection{E-type part of the Ward identities
and determination of the effective action}

Our next task is the analysis of the Ward identity (\ref{expward}).
  By substituting the
expansions (\ref{effexp4}), (\ref{expstr2}), (\ref{expstth2}) and the
results (\ref{efforder2}), (\ref{susytreer}),
(\ref{susytreeth}) obtained at order 2, and subsequently classifying terms 
 by the number of $\theta$'s,  the Ward identity can be split
into the following five equations: \\
$\calO (\del^4 \theta)$:
\begin{align}
\int \d\tau \left(
 - \Omega^{\del^2 \theta}_{m \alpha} \epsilon_\alpha a^m 
 + T^{\del^3}_{\alpha \beta} \epsilon_\beta \dot \theta_\alpha
- i (\epsilon \gamma^m \theta) 
  \frac{\delta \Gamma^{\del^4}}{\delta r^m}
+i (\vslash \epsilon)_\alpha 
  \frac{\delta \Gamma^{\del^3 \theta^2}}{\delta \theta_\alpha} \right)
= 0 \ , 
\label{wardp4th1}
\end{align}
$\calO (\del^3 \theta^3)$:
\begin{align}
\int \d\tau \left(
 - \Omega^{\del \theta^3}_{m \alpha} \epsilon_\alpha a^m 
 + T^{\del^2 \theta^2}_{\alpha \beta} \epsilon_\beta \dot \theta_\alpha
- i (\epsilon \gamma^m \theta) 
  \frac{\delta \Gamma^{\del^3 \theta^2}}{\delta r^m}
+i (\vslash \epsilon)_\alpha 
  \frac{\delta \Gamma^{\del^2 \theta^4}}{\delta \theta_\alpha} \right) 
= 0 \ ,
 \label{wardp3th3}
\end{align}
$\calO (\del^2 \theta^5)$:
\begin{align}
\int \d\tau \left(
 - \Omega^{\theta^5}_{m \alpha} \epsilon_\alpha a^m 
 + T^{\del \theta^4}_{\alpha \beta} \epsilon_\beta \dot \theta_\alpha
- i (\epsilon \gamma^m \theta) 
  \frac{\delta \Gamma^{\del^2 \theta^4}}{\delta r^m}
+i (\vslash \epsilon)_\alpha 
  \frac{\delta \Gamma^{\del \theta^6}}{\delta \theta_\alpha} \right) 
= 0 \ , 
\label{wardp2th5}
\end{align}
$\calO (\del \theta^7)$:
\begin{align}
\int \d\tau \left(
  T^{\theta^6}_{\alpha \beta} \epsilon_\beta \dot \theta_\alpha
- i (\epsilon \gamma^m \theta) 
  \frac{\delta \Gamma^{\del \theta^6}}{\delta r^m}
+i (\vslash \epsilon)_\alpha 
  \frac{\delta \Gamma^{\theta^8}}{\delta \theta_\alpha} \right)
= 0 \ , 
\label{wardp1th7}
\end{align}
$\calO (\theta^9)$:
\begin{align}
\int \d\tau \left(
- i (\epsilon \gamma^m \theta) 
  \frac{\delta \Gamma^{\theta^8}}{\delta r^m} \right)
= 0 \period
\label{wardp0th9}
\end{align}
By the E-type - D-type separation procedure explained previously,
 we decompose each of these equations into 3 types of local equations 
 (\ref{e0}) $\sim $ (\ref{dd2}). In the rest of this subsection, 
 we solve the purely E-type 
 equations of the type (\ref{e0}) to determine the coefficient 
 functions $f_i$'s. The other two types of equations, 
which will fix the SUSY transformation laws, will be studied 
 later.

\subsubsection{Analysis at $\calO(\p^4\epsilon\theta)$}
We begin our analysis by looking at the part with one power of 
 $\theta_\al$, \ie at $\calO(\p^4\epsilon\theta)$. 
The relevant E-type terms are produced by the last two terms in 
 the Ward identity (\ref{wardp4th1}). When we substitute the 
 explicit form of the effective action (\ref{eff4p4th0})
 and (\ref{eff4p3th2}), the resultant E-type terms turned out to be 
 not all independent. Thus we have to add appropriate total derivative 
 terms. In this way, we obtain 
 \begin{align}
- i (\epsilon \gamma^m \theta) 
  \frac{\delta \Gamma^{\del^4}}{\delta r^m}
&
+i (\vslash \epsilon)_\alpha 
  \frac{\delta \Gamma^{\del^3 \theta^2}}{\delta \theta_\alpha}
\simeq  \nn \\
&  -i 
\left(
 -2\,f^{{\p}^3{\theta}^2}\underscore1
   +\frac{1}{r}\frac{\d f^{{\p}^4}\underscore1}{\d r}
\right)\,v^4\,{r_i}\,
     (\epsilon {{\gamma }^i}\theta ) 
    + 2\,i\,G^{{\p}^3{\theta}^2}\underscore1\,v^2\,{a_i}\,
     (\epsilon {{\gamma }^i}\theta ) \brkeq  +
    4\,i\,G^{{\p}^3{\theta}^2}\underscore1\,{v_i}\,
     (v \cdot a)\,(\epsilon {{\gamma }^i}\theta ) 
   - i 
\left( 4\,f^{{\p}^4}\underscore1 - 2\,G^{{\p}^3{\theta}^2}\underscore1 \right)
\,v^2\,{v_i}\,
     (\epsilon {{\gamma }^i}{\dot{\theta}})
\label{etypep4th1}\comma 
\end{align}
where $G_1^{\del^3\theta^2}(r)$ is given by 
\begin{align}
G_1^{\del^3\theta^2}(r) \equiv \int^r r' f_1^{\del^3\theta^2} (r')\d r' 
\period
\end{align}
Although it is expressed as an integral, it is actually a local 
 expression. On the RHS of (\ref{etypep4th1}), the first term is 
 the only E-type term and hence it must vanish by itself. 
This gives the relation 
\begin{align}
{{f^{{\p}^3{\theta}^2}\underscore1} = 
    {\frac{1}{2\,r}\frac{\d f^{{\p}^4}\underscore1}{\d r}}} 
\label{diffeqth01} \comma 
\end{align}
which gives a direct connection 
between  $\Ga^{\del^4}$  and $\Ga^{\del^3\theta^2}$. 

\subsubsection{Analysis at $\calO(\p^3\epsilon\theta^3)$}

In an entirely similar manner,  E-type terms at this order 
 are produced by the last two terms of (\ref{wardp3th3}), and 
 after adding appropriate total derivatives, 
we arrive at the 
following expression consisting of 
3 independent E-type terms and 7 independent D-type terms, 
\begin{align}
& - i (\epsilon \gamma^m \theta) 
  \frac{\delta \Gamma^{\del^3 \theta^2}}{\delta r^m}
+
i (\vslash \epsilon)_\alpha 
  \frac{\delta \Gamma^{\del^2 \theta^4}}{\delta \theta_\alpha} 
\simeq  \nn \\
& 
 +4\,i\,v^2\,f^{{\p}^2{\theta}^4}\underscore1\,{r_j}\,{r_k}\,
   {v_l}\,(\epsilon {{\gamma }^{ikl}}\theta )\,
   (\theta {{\gamma }^{ij}}\theta ) - i
  v^2\,\left( 4\,f^{{\p}^2{\theta}^4}\underscore2 - 
     f^{{\p}^3{\theta}^2}\underscore1 - 
     4\,G^{{\p}^2{\theta}^4}\underscore1 \right) \,{v_j}\,
   (\epsilon {{\gamma }^{i}}\theta )\,
   (\theta {{\gamma }^{ij}}\theta ) \brkeq 
 - i 
  v^2\,{r_i}\,{r_j}\,{v_k}\,
   \left( 4\,f^{{\p}^2{\theta}^4}\underscore1 + 
     4\,f^{{\p}^2{\theta}^4}\underscore3 - 
     \frac{1}{r}\frac{\d f^{{\p}^3{\theta}^2}\underscore1}
       {\d r} \right) \,(\epsilon {{\gamma }^{i}}\theta )\,
   (\theta {{\gamma }^{jk}}\theta )  \nn \\
& +8\,i\,G^{{\p}^2{\theta}^4}\underscore1\,{r_j}\,(v \cdot a)\,
   (\epsilon {{\gamma }^{i}}\theta )\,
   (\theta {{\gamma }^{ij}}\theta ) - 
  4 i \,G^{{\p}^2{\theta}^4}\underscore3\,{r_j}\,{v_k}\,
   {a_i}\,(\epsilon {{\gamma }^{i}}\theta )\,
   (\theta {{\gamma }^{jk}}\theta ) \brkeq 
- 
  4 i \,G^{{\p}^2{\theta}^4}\underscore3\,{r_j}\,{v_i}\,
   {a_k}\,(\epsilon {{\gamma }^{i}}\theta )\,
   (\theta {{\gamma }^{jk}}\theta ) - i  
  v^2\,\left( f^{{\p}^3{\theta}^2}\underscore1 - 
     4\,G^{{\p}^2{\theta}^4}\underscore1 \right) \,{r_j}\,
   (\epsilon {{\gamma }^{i}}{\dot{\theta}})\,
   (\theta {{\gamma }^{ij}}\theta ) \brkeq 
 - i 
  \left( -2\,f^{{\p}^3{\theta}^2}\underscore1 
+ 
     4\,G^{{\p}^2{\theta}^4}\underscore3 \right) \,{r_j}\,
   {v_i}\,{v_k}\,(\epsilon {{\gamma }^{i}}{\dot{\theta}})\,
   (\theta {{\gamma }^{jk}}\theta ) 
+
  8\,i\,v^2\,G^{{\p}^2{\theta}^4}\underscore1\,{r_j}\,
   (\epsilon {{\gamma }^{i}}\theta )\,
   ({\dot{\theta}}{{\gamma }^{ij}}\theta ) \brkeq  -  
  8\,i\,G^{{\p}^2{\theta}^4}\underscore3\,{r_j}\,{v_i}\,
   {v_k}\,(\epsilon {{\gamma }^{i}}\theta )\,
   ({\dot{\theta}}{{\gamma }^{jk}}\theta ) 
\label{etypeth3}, 
\end{align}
where we have defined 
\begin{align}
  G^{\p^2 \theta^4}_i(r)
\equiv \int^r \!\!  r' f_i^{\p^2 \theta^4}(r') \, \d r' , \ \ \ 
  G^{\p \theta^6}_i(r)
\equiv \int^r \!\!  r' f_i^{\p \theta^6}(r') \, \d r'  .
\end{align}
It can be checked that the E-type structures, the first 3 terms 
 on the RHS, cannot be related by Fierz identities and are 
 independent. Setting them separately to zero, we obtain 
\begin{align}
 f^{{\p}^2{\theta}^4}\underscore1 & = 0  
\label{p2th410}, \\
  f^{{\p}^2{\theta}^4}\underscore2 & =  
   \frac{f^{{\p}^3{\theta}^2}\underscore1}{4} , 
\label{diffeqth21} \\  
  f^{{\p}^2{\theta}^4}\underscore3 & = 
   \frac{1}{4\,r}\frac{\d 
       f^{{\p}^3{\theta}^2}\underscore1}{\d r}  
\label{diffeqth22} . 
\end{align}
Evidently, these relations determine $\Ga^{\del^2 \theta^4}$ in terms 
 of $f_1^{\del^3 \theta^2}$, which in turn has already been 
 related to $f_1^{\del^4}$. 
\subsubsection{Analysis at $\calO(\p^2\epsilon\theta^5)$}

Beginning at this order with 6 spinors, our task becomes much more
difficult, since, in addition to adding total derivatives, 
we must find and apply 
 judicious Fierz identities 
in order to bring the relevant E-type terms to completely 
independent expressions. The last two terms of the Ward 
 identity  (\ref{wardp2th5}) yields
\begin{align}
& - i (\epsilon \gamma^m \theta) 
  \frac{\delta \Gamma^{\del^2 \theta^4}}{\delta r^m}
+
i (\vslash \epsilon)_\alpha 
  \frac{\delta \Gamma^{\del \theta^6}}{\delta \theta_\alpha} 
\simeq  \nn \\
& -2\,f^{\p{\theta}^6}\underscore2\,{r_{{i_1}}}\,
   {r_{{i_2}}}\,{v_{{i_3}}}\,(r \cdot v)\,
   (\epsilon {{\gamma }^{{i_3}}}\theta )\,
   (\theta {{\gamma }^{{a_1}{i_1}}}\theta )\,
   (\theta {{\gamma }^{{a_1}{i_2}}}\theta ) 
+   2\,f^{\p{\theta}^6}\underscore1\,{r_{{i_1}}}\,
   {v_{{i_2}}}\,{v_{{i_3}}}\,
   (\epsilon {{\gamma }^{{i_2}}}\theta )\,
   (\theta {{\gamma }^{{a_1}{i_1}}}\theta )\,
   (\theta {{\gamma }^{{a_1}{i_3}}}\theta ) \brkeq + 
  2\,v^2\,f^{\p{\theta}^6}\underscore1  \,
   {r_{{i_1}}}\,(\epsilon {{\gamma }^{{a_1}}}\theta )\,
   (\theta {{\gamma }^{{a_1}{a_2}}}\theta )\,
   (\theta {{\gamma }^{{a_2}{i_1}}}\theta )  - 
  2\,f^{\p{\theta}^6}\underscore1\,{v_{{i_1}}}\,
   (r \cdot v)\,(\epsilon {{\gamma }^{{a_1}}}\theta )\,
   (\theta {{\gamma }^{{a_1}{a_2}}}\theta )\,
   (\theta {{\gamma }^{{a_2}{i_1}}}\theta )  \brkeq - 
  2\,f^{{\p}^2{\theta}^4}\underscore2\,{v_{{i_1}}}\,
   (\epsilon {{\gamma }^{{a_1}}}{\dot{\theta}})\,
   (\theta {{\gamma }^{{a_1}{a_2}}}\theta )\,
   (\theta {{\gamma }^{{a_2}{i_1}}}\theta )  - 
  2\,f^{\p{\theta}^6}\underscore1\,{r_{{i_1}}}\,
   {v_{{i_2}}}\,{v_{{i_3}}}\,
   (\epsilon {{\gamma }^{{a_1}{a_2}{i_2}}}\theta )\,
   (\theta {{\gamma }^{{a_1}{i_3}}}\theta )\,
   (\theta {{\gamma }^{{a_2}{i_1}}}\theta ) \brkeq + 
  2\,f^{\p{\theta}^6}\underscore1\,{r_{{i_1}}}\,
   {v_{{i_2}}}\,{v_{{i_3}}}\,
   (\epsilon {{\gamma }^{{a_1}{i_1}{i_2}}}\theta )\,
   (\theta {{\gamma }^{{a_1}{a_2}}}\theta )\,
   (\theta {{\gamma }^{{a_2}{i_3}}}\theta )  
+ 
  4\,f^{\p{\theta}^6}\underscore2\,{r_{{i_1}}}\,
   {r_{{i_2}}}\,{v_{{i_3}}}\,(r \cdot v)\,
   (\epsilon {{\gamma }^{{a_1}}}\theta )\,
   (\theta {{\gamma }^{{a_1}{i_1}}}\theta )\,
   (\theta {{\gamma }^{{i_2}{i_3}}}\theta )  \brkeq- 
  2\,f^{{\p}^2{\theta}^4}\underscore3\,{r_{{i_1}}}\,
   {r_{{i_2}}}\,{v_{{i_3}}}\,
   (\epsilon {{\gamma }^{{a_1}}}{\dot{\theta}})\,
   (\theta {{\gamma }^{{a_1}{i_1}}}\theta )\,
   (\theta {{\gamma }^{{i_2}{i_3}}}\theta )   + 
2\,f^{\p{\theta}^6}\underscore2 \, 
  v^2\,{r_{{i_1}}}\,{r_{{i_2}}}\,{r_{{i_3}}}\,
   (\epsilon {{\gamma }^{{i_1}}}\theta )\,
   (\theta {{\gamma }^{{a_1}{i_2}}}\theta )\,
   (\theta {{\gamma }^{{a_1}{i_3}}}\theta )
\brkeq  
+ 
  \left( -2\,f^{\p{\theta}^6}\underscore1 + 
     2\,f^{{\p}^2{\theta}^4}\underscore3 \right) \,
   {r_{{i_1}}}\,{v_{{i_2}}}\,{v_{{i_3}}}\,
   (\epsilon {{\gamma }^{{a_1}}}\theta )\,
   (\theta {{\gamma }^{{a_1}{i_2}}}\theta )\,
   (\theta {{\gamma }^{{i_1}{i_3}}}\theta ) \brkeq 
+    \left( -2\,f^{\p{\theta}^6}\underscore1 
 + 
     \frac{1}{r}\frac{\d f^{{\p}^2{\theta}^4}\underscore2}
       {\d r} \right)  
  {r_{{i_1}}}\,{v_{{i_2}}}\,{v_{{i_3}}}\,
   (\epsilon {{\gamma }^{{i_1}}}\theta )\,
   (\theta {{\gamma }^{{a_1}{i_2}}}\theta )\,
   (\theta {{\gamma }^{{a_1}{i_3}}}\theta )\,
\brkeq 
+    \left( 4\,f^{\p{\theta}^6}\underscore2 + 
     \frac{1}{r}\frac{\d f^{{\p}^2{\theta}^4}\underscore3}
       {\d r} \right)
  {r_{{i_1}}}\,{r_{{i_2}}}\,{r_{{i_3}}}\,{v_{{i_4}}}\,
   {v_{{i_5}}}\,(\epsilon {{\gamma }^{{i_1}}}\theta )\,
   (\theta {{\gamma }^{{i_2}{i_4}}}\theta )\,
   (\theta {{\gamma }^{{i_3}{i_5}}}\theta )\,
\brkeq - 
  4\,f^{\p{\theta}^6}\underscore2\,{r_{{i_1}}}\,
   {r_{{i_2}}}\,{r_{{i_3}}}\,{v_{{i_4}}}\,{v_{{i_5}}}\,
   (\epsilon {{\gamma }^{{a_1}{i_1}{i_4}}}\theta )\,
   (\theta {{\gamma }^{{a_1}{i_2}}}\theta )\,
   (\theta {{\gamma }^{{i_3}{i_5}}}\theta ) 
\label{sv2th7} , 
\end{align}
where we have set $f_1^{\del^2\theta^4}$ to zero according to (\ref{p2th410}). 
The E-type terms in this expression are not (even algebraically) independent
and we must make use of various Fierz identities\footnote{These Fierz 
identities, many of which are quite complicated,  are generated using an
efficient algorithm described in the Appendix A of
\cite{Kaz-Mura2}.  In particular, the ones involving several different 
spinors and/or  with large number of free-indices (\ie uncontracted indices)
can be extremely complicated.
For example, the longest five-free-index identity consists of
109 terms.} 
as well as integration by parts to reduce them to independent
 forms. 
Since these Fierz identities and
the results generated by their applications at intermediate steps are
 too space-filling to be displayed here, we
shall only sketch the reduction procedure. 

First, we rewrite the last term on the RHS of  (\ref{sv2th7})
 by the use of several 
 5-free-index type Fierz identities.  The results so obtained are
 further reduced by using the 3-free-index Fierz identities of 
 the following form:
\begin{align}
(\epsilon {{\gamma }^{{a_1}i}}\theta )\,
     (\theta {{\gamma }^{{a_1}jk}}\theta )  = 
    \and +(\epsilon {{\gamma }^{{a_1}jk}}\theta )\,
       (\theta {{\gamma }^{{a_1}i}}\theta )-2
      (\epsilon {{\gamma }^k}\theta )\,
       (\theta {{\gamma }^{ij}}\theta )+2
      (\epsilon {{\gamma }^j}\theta )\,
       (\theta {{\gamma }^{ik}}\theta ) \brkeq 
     +2(\epsilon {{\gamma }^i}\theta )\,
       (\theta {{\gamma }^{jk}}\theta )-2
      (\epsilon \theta )\,
       (\theta {{\gamma }^{ijk}}\theta )-
      (\epsilon {{\gamma }^{{a_1}}}\theta )\,
       (\theta {{\gamma }^{{a_1}k}}\theta )\,
       {{\delta }_{ij}} \brkeq 
     +(\epsilon {{\gamma }^{{a_1}}}\theta )\,
       (\theta {{\gamma }^{{a_1}j}}\theta )\,
       {{\delta }_{ik}},
   \landfz (\epsilon {{\gamma }^{{a_1}ij}}\theta )\,
     (\theta {{\gamma }^{{a_1}k}}\theta )  = 
    \and -(\epsilon {{\gamma }^{{a_1}jk}}\theta )\,
       (\theta {{\gamma }^{{a_1}i}}\theta )+
      (\epsilon {{\gamma }^{{a_1}ik}}\theta )\,
       (\theta {{\gamma }^{{a_1}j}}\theta )+
      (\epsilon {{\gamma }^k}\theta )\,
       (\theta {{\gamma }^{ij}}\theta ) \brkeq 
     -(\epsilon {{\gamma }^j}\theta )\,
       (\theta {{\gamma }^{ik}}\theta )+
      (\epsilon {{\gamma }^i}\theta )\,
       (\theta {{\gamma }^{jk}}\theta )+3
      (\epsilon \theta )\,
       (\theta {{\gamma }^{ijk}}\theta )  , 
   \landfz (\theta {{\gamma }^{{a_1}j}}\theta )\,
     (\theta {{\gamma }^{{a_1}ik}}\theta )  = 
    \and  (\theta {{\gamma }^{{a_1}k}}\theta )\,
       (\theta {{\gamma }^{{a_1}ij}}\theta )+
      (\theta {{\gamma }^{{a_1}i}}\theta )\,
       (\theta {{\gamma }^{{a_1}jk}}\theta ) .
\end{align} 
At this stage, the resultant terms become {\it algebraically} independent.
To make them truly independent in the sense defined before, we must 
 add total derivative terms. In this way, we finally obtain 
 the following completely independent form for the E-type terms:
\begin{align}
\Bigg(
- i (\epsilon &  \gamma^m \theta)  
  \frac{\delta \Gamma^{\del^2 \theta^4}}{\delta r^m}
+i (\vslash \epsilon)_\alpha 
  \frac{\delta \Gamma^{\del \theta^6}}{\delta \theta_\alpha} \Bigg)
\Bigg|_{\text{E-type term}} \simeq \nn \\
& \left( 4\,f^{\p{\theta}^6}\underscore1 - 
     \frac{4\,r^2\,f^{\p{\theta}^6}\underscore2}{5} - 
     \frac{4\,G^{\p{\theta}^6}\underscore2}{5} \right) \,
   {r_{{i_1}}}\,{v_{{i_2}}}\,{v_{{i_3}}}\,
   (\epsilon {{\gamma }^{{i_2}}}\theta )\,
   (\theta {{\gamma }^{{a_1}{i_1}}}\theta )\,
   (\theta {{\gamma }^{{a_1}{i_3}}}\theta ) \brkeq 
+ 
  \left( 2\,f^{\p{\theta}^6}\underscore1 - 
     \frac{2\,r^2\,f^{\p{\theta}^6}\underscore2}{5} - 
     \frac{2\,G^{\p{\theta}^6}\underscore2}{5} \right) v^2 \,
   {r_{{i_1}}}\,(\epsilon {{\gamma }^{{a_1}}}\theta )\,
   (\theta {{\gamma }^{{a_1}{a_2}}}\theta )\,
   (\theta {{\gamma }^{{a_2}{i_1}}}\theta ) \brkeq 
+ 
  \left( 4\,f^{\p{\theta}^6}\underscore1 - 
     \frac{4\,r^2\,f^{\p{\theta}^6}\underscore2}{5} - 
     \frac{4\,G^{\p{\theta}^6}\underscore2}{5} \right) \,
   {r_{{i_1}}}\,{v_{{i_2}}}\,{v_{{i_3}}}\,
   (\epsilon {{\gamma }^{{a_1}{a_2}{i_2}}}\theta )\,
   (\theta {{\gamma }^{{a_1}{i_1}}}\theta )\,
   (\theta {{\gamma }^{{a_2}{i_3}}}\theta ) \brkeq 
+ 
  \left( -4\,r^2\,f^{\p{\theta}^6}\underscore2 + 
     2\,f^{{\p}^2{\theta}^4}\underscore3 - 
     8\,G^{\p{\theta}^6}\underscore2 \right) \,
   {r_{{i_1}}}\,{v_{{i_2}}}\,{v_{{i_3}}}\,
   (\epsilon {{\gamma }^{{a_1}}}\theta )\,
   (\theta {{\gamma }^{{a_1}{i_2}}}\theta )\,
   (\theta {{\gamma }^{{i_1}{i_3}}}\theta ) \brkeq 
+ 
  \left( 6\,f^{\p{\theta}^6}\underscore1 - 
     \frac{6\,r^2\,f^{\p{\theta}^6}\underscore2}{5} - 
     \frac{6\,G^{\p{\theta}^6}\underscore2}{5} \right) \,
   {r_{{i_1}}}\,{v_{{i_2}}}\,{v_{{i_3}}}\,
   (\epsilon \theta )\,(\theta {{\gamma }^{{a_1}{i_2}}}
    \theta )\,(\theta {{\gamma }^{{a_1}{i_1}{i_3}}}\theta 
    ) \brkeq 
+   \left( -4\,f^{\p{\theta}^6}\underscore1 - 
     \frac{6\,r^2\,f^{\p{\theta}^6}\underscore2}{5} - 
     \frac{16\,G^{\p{\theta}^6}\underscore2}{5} + 
     \frac{1}{r}\frac{\d f^{{\p}^2{\theta}^4}\underscore2}
       {\d r} \right)  
 {r_{{i_1}}}\,{v_{{i_2}}}\,{v_{{i_3}}}\,
   (\epsilon {{\gamma }^{{i_1}}}\theta )\,
   (\theta {{\gamma }^{{a_1}{i_2}}}\theta )\,
   (\theta {{\gamma }^{{a_1}{i_3}}}\theta )\,
\brkeq 
+    \left( 18\,f^{\p{\theta}^6}\underscore2 + 
     \frac{1}{r}\frac{\d f^{{\p}^2{\theta}^4}\underscore3}
       {\d r} \right) \,
  {r_{{i_1}}}\,{r_{{i_2}}}\,{r_{{i_3}}}\,{v_{{i_4}}}\,
   {v_{{i_5}}}\,(\epsilon {{\gamma }^{{i_1}}}\theta )\,
   (\theta {{\gamma }^{{i_2}{i_4}}}\theta )\,
   (\theta {{\gamma }^{{i_3}{i_5}}}\theta ) . 
\end{align}
Setting the coefficient of each term to zero and making some rearrangements, 
 we obtain the relations
\begin{align}
f^{\p{\theta}^6}\underscore2 &= 
   -\frac{1}{18\,r}\frac{\d f^{{\p}^2{\theta}^4}\underscore3}
     {\d r}  \label{diffeqth41}, \\
f^{\p{\theta}^6}\underscore1  = 
   \frac{G^{\p{\theta}^6}\underscore2}{5} - 
    \frac{r}{90}\frac{\d f^{{\p}^2{\theta}^4}\underscore3}
      {\d r}  ,  \ \ 
f^{{\p}^2{\theta}^4}\underscore3  &= 
   4\,G^{\p{\theta}^6}\underscore2 - 
   \frac{r}{9} \frac{\d f^{{\p}^2{\theta}^4}\underscore3}{\d r}  , \ \
  \frac{\d f^{{\p}^2{\theta}^4}\underscore2}{\d r}  = 
   4\,r\,G^{\p{\theta}^6}\underscore2 - 
    \frac{r^2}{9} \frac{\d f^{{\p}^2{\theta}^4}\underscore3}{\d r}.
\end{align}
The first of these relations, (\ref{diffeqth41}), 
 coincides with the one previously 
obtained in \cite{Hyunetal} 
in the eikonal approximation. The other 3 relations are new. 
Although we shall not elaborate on it, they can be used to 
fix the dependence on $r$ of various coefficients without resort
 to the analysis of terms with higher number of $\theta$'s. 

\subsubsection{Analysis at $\calO(\p \epsilon\theta^7)$}

We now come to the structure with 8 spinors.  The relevant 
 E-type terms are produced by the second and the third terms 
of the Ward identity (\ref{wardp1th7}). The procedure for 
 reducing these terms to independent ones parallels the one 
 at $\calO(\p^2\epsilon\theta^5)$. Uses of intricate five-free-index  and 
 other types of Fierz identities together with integration by parts 
 leads to the following independent E-type terms:
\begin{align}
& \Bigg(
- i (\epsilon \gamma^m \theta)  
  \frac{\delta \Gamma^{\del \theta^6}}{\delta r^m}
+i (\vslash \epsilon)_\alpha 
  \frac{\delta \Gamma^{\theta^8}}{\delta \theta_\alpha} \Bigg)
\Bigg|_{\text{E-type term}} \simeq \nn \\
& \left( \frac{-24\,r^2\,f^{{\theta}^8}\underscore3}{5} + 
     f^{\p{\theta}^6}\underscore2 - 
     \frac{32\,G^{{\theta}^8}\underscore3}{5} \right) \,
   {r_{{i_1}}}\,{r_{{i_2}}}\,{v_{{i_3}}}\,
   (\epsilon {{\gamma }^{{a_1}}}\theta )\,
   (\theta {{\gamma }^{{a_1}{i_3}}}\theta )\,
   (\theta {{\gamma }^{{a_2}{i_1}}}\theta )\,
   (\theta {{\gamma }^{{a_2}{i_2}}}\theta )  \brkeq
+ 
  \left( -4\,f^{{\theta}^8}\underscore2 - 
     \frac{192\,r^2\,f^{{\theta}^8}\underscore3}{25} - 
     \frac{1}{r}\frac{\d f^{\p{\theta}^6}\underscore1}{\d r}
 - \frac{96\,G^{{\theta}^8}\underscore3}{25}
     \right) \,{r_{{i_1}}}\,{r_{{i_2}}}\,{v_{{i_3}}}\,
   (\epsilon {{\gamma }^{{i_1}}}\theta )\,
   (\theta {{\gamma }^{{a_1}{a_2}}}\theta )\,
   (\theta {{\gamma }^{{a_1}{i_2}}}\theta )\,
   (\theta {{\gamma }^{{a_2}{i_3}}}\theta ) \brkeq
+ 
  \left( -4\,f^{{\theta}^8}\underscore2 - 
     \frac{8\,r^2\,f^{{\theta}^8}\underscore3}{5} - 
     \frac{184\,G^{{\theta}^8}\underscore3}{5} \right) \,
   {r_{{i_1}}}\,{r_{{i_2}}}\,{v_{{i_3}}}\,
   (\epsilon {{\gamma }^{{a_1}}}\theta )\,
   (\theta {{\gamma }^{{a_1}{i_1}}}\theta )\,
   (\theta {{\gamma }^{{a_2}{i_2}}}\theta )\,
   (\theta {{\gamma }^{{a_2}{i_3}}}\theta ) \brkeq + 
  \left( -8\,f^{{\theta}^8}\underscore1 + 
     \frac{4\,r^4\,f^{{\theta}^8}\underscore3}{25} + 
     \frac{12\,r^2\,G^{{\theta}^8}\underscore3}{25} + 
     \frac{12\,\widetilde{G}^{{\theta}^8}\underscore3}{25} + 
     \frac{4\,\overline{G}H^{{\theta}^8}\underscore3}{25} \right) \nn \\
& \hspace{7cm} \times \,
   {v_{{i_1}}}\,(\epsilon {{\gamma }^{{a_1}{a_2}{i_1}}}
    \theta )\,(\theta {{\gamma }^{{a_1}{a_3}}}\theta )\,
   (\theta {{\gamma }^{{a_2}{a_4}}}\theta )\,
   (\theta {{\gamma }^{{a_3}{a_4}}}\theta )  \brkeq + 
  \left( -16\,f^{{\theta}^8}\underscore1 - 
     \frac{24\,r^4\,f^{{\theta}^8}\underscore3}{25} + 
     f^{\p{\theta}^6}\underscore1 - 
     4\,G^{{\theta}^8}\underscore2 - 
     \frac{72\,r^2\,G^{{\theta}^8}\underscore3}{25} - 
     \frac{72\,\widetilde{G}^{{\theta}^8}\underscore3}{25} - 
     \frac{24\,\overline{G}^{{\theta}^8}\underscore3}{25} \right)  \nn \\
& \hspace{7cm} \times \,
   {v_{{i_1}}}\,(\epsilon {{\gamma }^{{a_1}}}\theta )\,
   (\theta {{\gamma }^{{a_1}{a_2}}}\theta )\,
   (\theta {{\gamma }^{{a_2}{a_3}}}\theta )\,
   (\theta {{\gamma }^{{a_3}{i_1}}}\theta ) \brkeq + 
  \left( -4\,f^{{\theta}^8}\underscore2 + 
     \frac{32\,r^2\,f^{{\theta}^8}\underscore3}{25} + 
     \frac{16\,G^{{\theta}^8}\underscore3}{25} \right) \,
   {r_{{i_1}}}\,{r_{{i_2}}}\,{v_{{i_3}}}\,
   (\epsilon {{\gamma }^{{a_1}{i_1}{i_3}}}\theta )\,
   (\theta {{\gamma }^{{a_1}{a_2}}}\theta )\,
   (\theta {{\gamma }^{{a_2}{a_3}}}\theta )\,
   (\theta {{\gamma }^{{a_3}{i_2}}}\theta )  \brkeq + 
  \left( -4\,f^{{\theta}^8}\underscore2 + 
     \frac{8\,r^2\,f^{{\theta}^8}\underscore3}{5} + 
     \frac{24\,G^{{\theta}^8}\underscore3}{5} \right) \,
   {r_{{i_1}}}\,{r_{{i_2}}}\,{v_{{i_3}}}\,
   (\epsilon {{\gamma }^{{a_1}{a_2}{i_3}}}\theta )\,
   (\theta {{\gamma }^{{a_1}{i_1}}}\theta )\,
   (\theta {{\gamma }^{{a_2}{a_3}}}\theta )\,
   (\theta {{\gamma }^{{a_3}{i_2}}}\theta )  \brkeq + 
  \left( 4\,f^{{\theta}^8}\underscore2 + 
     8\,r^2\,f^{{\theta}^8}\underscore3 - 
     2\,f^{\p{\theta}^6}\underscore2 + 
     8\,G^{{\theta}^8}\underscore3 \right) \,{r_{{i_1}}}\,
   {r_{{i_2}}}\,{v_{{i_3}}}\,
   (\epsilon {{\gamma }^{{a_1}}}\theta )\,
   (\theta {{\gamma }^{{a_1}{a_2}}}\theta )\,
   (\theta {{\gamma }^{{a_2}{i_1}}}\theta )\,
   (\theta {{\gamma }^{{i_2}{i_3}}}\theta ) \brkeq + 
  \left( 56\,f^{{\theta}^8}\underscore3 + 
     \frac{1}
      {r} \frac{\d f^{\p{\theta}^6}\underscore2}{\d r}
 \right) \,{r_{{i_1}}}\,{r_{{i_2}}}\,
   {r_{{i_3}}}\,{r_{{i_4}}}\,{v_{{i_5}}}\,
   (\epsilon {{\gamma }^{{i_1}}}\theta )\,
   (\theta {{\gamma }^{{a_1}{i_2}}}\theta )\,
   (\theta {{\gamma }^{{a_1}{i_3}}}\theta )\,
   (\theta {{\gamma }^{{i_4}{i_5}}}\theta ) ,
\end{align}
where 
\begin{align}
 G_i^{\theta^8}(r) \equiv \int^r \!\! r' f_i^{\theta^8}(r') \, \d r' , \ \ \ 
 \widetilde{G}_i^{\theta^8}(r) \equiv \int^r \!\! r' G_i^{\theta^8}(r') \, \d r' , \ \ \
 \overline{G}_i^{\theta^8}(r) \equiv \int^r \!\! r^{\prime 3} f_i^{\theta^8}(r') \, \d r' . 
\end{align}
Setting the coefficients  to zero and rearranging the resultant 
equations, we get
\begin{align}
 f^{\p{\theta}^6}\underscore2 & = 
   \frac{56\,r^2\,f^{{\theta}^8}\underscore3}{13} 
\label{diffeqth61}  , \\
  \frac{\d f^{\p{\theta}^6}\underscore1}{\d r} & = 
   \frac{-112\,r^3\,f^{{\theta}^8}\underscore3}{13} 
\label{diffeqth62}  ,  
\end{align} 
\begin{align}
\frac{\d f^{\p{\theta}^6}\underscore2}{\d r} & = 
   -56\,r\,f^{{\theta}^8}\underscore3 ,  \ \ 
  G^{{\theta}^8}\underscore3  =  -
   \frac{r^2\,f^{{\theta}^8}\underscore3}
    {13} ,  \ \ 
 f^{{\theta}^8}\underscore1  = 
   \frac{10\,r^4\,f^{{\theta}^8}\underscore3 + 
      39\,\widetilde{G}^{{\theta}^8}\underscore3 + 
      13\,\overline{G}^{{\theta}^8}\underscore3}{650}  , \\ 
  f^{{\theta}^8}\underscore2 & = 
   \frac{4\,r^2\,f^{{\theta}^8}\underscore3}{13} ,  \ \
  f^{\p{\theta}^6}\underscore1  = 
   \frac{4\,\left( 80\,r^4\,f^{{\theta}^8}\underscore3 + 
        325\,G^{{\theta}^8}\underscore2 + 
        312\,\widetilde{G}^{{\theta}^8}\underscore3 + 
        104\,\overline{G}^{{\theta}^8}\underscore3 \right) }{325} , 
\end{align}

\subsubsection{Analysis at $\calO(\epsilon\theta^9)$}

Finally, we are left with the structures with 10 spinors.  
These structures without any derivatives 
have already been studied by Paban {\it et al.}\cite{Pabanetal1}. However, as 
they examined equations weaker than the actual invariance conditions, 
 our results will give stronger constraints\footnote{However,
  the extra solutions allowed by the  weaker conditions do not 
 satisfy physical requirements and their 
 effective action agrees with our result (\ref{resth8}).}. 
Starting from the Ward identity (\ref{wardp0th9}), we
 rewrite the terms with five free indecies by using five-free-index Fierz
identities.  Then the Ward identity can be brought to the form 
\begin{align}
& - i (\epsilon \gamma^m \theta) 
  \frac{\delta \Gamma^{\theta^8}}{\delta r^m} \simeq \nn \\ 
&   \left( 4\,\imag \,f^{{\theta}^8}\underscore3 + 
     \frac{4\,\imag }{15}\,r
      \frac{\d f^{{\theta}^8}\underscore3}{\d r} \right) 
{r_{{i_1}}}\,{r_{{i_2}}}\,{r_{{i_3}}}\,
   (\epsilon {{\gamma }^{{a_1}}}\theta )\,
   (\theta {{\gamma }^{{a_1}{a_2}}}\theta )\,
   (\theta {{\gamma }^{{a_2}{i_1}}}\theta )\,
   (\theta {{\gamma }^{{a_3}{i_2}}}\theta )\,
   (\theta {{\gamma }^{{a_3}{i_3}}}\theta )\,
\brkeq 
   + 
   \left( \frac{-\imag }{r}
      \frac{\d f^{{\theta}^8}\underscore2}{\d r}  + 
     \frac{4\,\imag }{15}\,r
      \frac{\d f^{{\theta}^8}\underscore3}{\d r} \right) 
{r_{{i_1}}}\,{r_{{i_2}}}\,{r_{{i_3}}}\,
   (\epsilon {{\gamma }^{{i_1}}}\theta )\,
   (\theta {{\gamma }^{{a_1}{a_2}}}\theta )\,
   (\theta {{\gamma }^{{a_1}{i_2}}}\theta )\,
   (\theta {{\gamma }^{{a_2}{a_3}}}\theta )\,
   (\theta {{\gamma }^{{a_3}{i_3}}}\theta )\,
\brkeq 
   +    \left( \frac{-\imag }{r}
      \frac{\d f^{{\theta}^8}\underscore1}{\d r} + 
     \frac{2\,\imag }{195}\,r^3
      \frac{\d f^{{\theta}^8}\underscore3}{\d r} \right) 
{r_{{i_1}}}\,(\epsilon {{\gamma }^{{i_1}}}\theta )\,
   (\theta {{\gamma }^{{a_1}{a_2}}}\theta )\,
   (\theta {{\gamma }^{{a_1}{a_3}}}\theta )\,
   (\theta {{\gamma }^{{a_2}{a_4}}}\theta )\,
   (\theta {{\gamma }^{{a_3}{a_4}}}\theta )\,
\brkeq   
+    \left( 2\,\imag \,f^{{\theta}^8}\underscore2 + 
     \frac{8\,\imag }{195}\,r^3
      \frac{\d f^{{\theta}^8}\underscore3}{\d r} \right)
{r_{{i_1}}}\,(\epsilon {{\gamma }^{{a_1}}}\theta )\,
   (\theta {{\gamma }^{{a_1}{a_2}}}\theta )\,
   (\theta {{\gamma }^{{a_2}{a_3}}}\theta )\,
   (\theta {{\gamma }^{{a_3}{a_4}}}\theta )\,
   (\theta {{\gamma }^{{a_4}{i_1}}}\theta ) .
\end{align}
Setting the coefficient of each term to zero and rearranging the
resultant 
equations,  we get,
\begin{align}
 f^{{\theta}^8}\underscore2  & = 
   \frac{-4\,r^3}{195}\frac{\d f^{{\theta}^8}\underscore3}{\d r} 
\label{diffeqth81},\\ 
  f^{{\theta}^8}\underscore3 & = 
   - \frac{r}{15}\frac{\d f^{{\theta}^8}\underscore3}{\d r} 
\label{fixth8}  ,\\ 
  \frac{\d f^{{\theta}^8}\underscore1}{\d r} & = 
   \frac{2\,r^4}{195} \frac{\d f^{{\theta}^8}\underscore3}{\d r} 
\label{diffeqth82} ,\\ 
  \frac{\d f^{{\theta}^8}\underscore2}{\d r} & = 
   \frac{4\,r^2}{15} \frac{\d f^{{\theta}^8}\underscore3}{\d r} .
\end{align}

\subsubsection{Determination of the effective action}

Having found all the relations imposed by the Ward identities, 
we now combine them to determine the coefficient functions $f_i$. 
First, by solving the differential equation (\ref{fixth8}), one finds 
\begin{align}
 f^{{\theta}^8}\underscore3 \propto \frac{1}{r^{15}}.
\end{align}
Now  we put this result into (\ref{diffeqth01}), (\ref{diffeqth21}),
(\ref{diffeqth22}), (\ref{diffeqth41}), 
(\ref{diffeqth61}), (\ref{diffeqth62}), (\ref{diffeqth81}) and 
(\ref{diffeqth82}), none of which contains integrated coefficients 
 $G_i$'s. There can be two choices of the physical 
boundary condition for solving these set of 
 differential equations, depending on one's view. One choice would be 
 to require that $\calL^{(4)}$ should be finite as $r\rightarrow \infty$. 
A slightly stronger alternative is that $\calL^{(4)}$  should 
  not only be finite but should vanish in the above limit. 
If we adopt the former, we get 
\begin{align}
& f^{{\p}^4}\underscore1 = b + \frac{c}{r^7} , 
\label{fp4}\\  
&  f^{{\p}^3{\theta}^2}\underscore1 = 
   \frac{-7\,c}{2\,r^9} ,\label{fp3th2} \\ 
&  f^{{\p}^2{\theta}^4}\underscore1 = 0 , \ \ 
  f^{{\p}^2{\theta}^4}\underscore2 = 
   \frac{-7\,c}{8\,r^9} , \ \ 
  f^{{\p}^2{\theta}^4}\underscore3 = 
   \frac{63\,c}{8\,r^{11}} , \label{fp2th4} \\ 
&  f^{\p{\theta}^6}\underscore1 = \frac{7\,c}{8\,r^{11}} , \ \ 
   f^{\p{\theta}^6}\underscore2 = 
   \frac{77\,c}{16\,r^{13}} ,  \\
&  f^{{\theta}^8}\underscore1 = 
\frac{c}{64\,r^{11}}
, \ \ 
  f^{{\theta}^8}\underscore2 = \frac{11\,c}{32\,r^{13}} , \ \  
   f^{{\theta}^8}\underscore3 = 
   \frac{143\,c}{128\,r^{15}} ,
\label{resth8} 
\end{align}
where $b$ and  $c$ are finite constants, while the latter stronger 
 condition sets $b$ to zero. At first sight, the presence of a term like 
$bv^4$ appears to violate the cluster property. This is certainly correct 
 if such a term is generated by some interactions of the underlying theory. 
Actually, in the case of Matrix theory a simple dimensional analysis 
tells us that the coefficient $b$ must be proportional to $g^{-14/3}$, 
 where $g$ is the gauge coupling constant, and hence could only be of 
non-perturbative origin.
However, since we do not make any assumption about the underlying theory, 
 one may simply accept such a term as describing
a self-interaction of a D-particle. As the rest of our analysis 
 is not affected by the presence of $b$, we will keep it. 

Let us briefly compare our results (\ref{fp4}) $\sim$ (\ref{resth8}) with 
 those obtained previously by various authors. We should distinguish 
 two categories:
\begin{itemize}
	\item General analysis without assuming underlying theory:\quad 
This type of analysis at order 4 
was initiated by Paban {\it et al.} \cite{Pabanetal1} for 
 the $\calO(\theta^8)$ part which does not contain any derivatives and 
 later extended by Hyun {\it et al.}\cite{Hyunetal}
 to the full structures containing 
 all the allowed powers of $\theta$. 
In spite of the fact that these analyses 
were incomplete in several senses, as already explained before, 
 our complete fully 
 off-shell results agree 
 precisely with those obtained in \cite{Hyunetal}.
This can be \lq explained'  by our method of 
E-type - D-type separation. Although careful analysis of independent basis 
 was crucial, the E-type part of the equations turned out to be 
 essentially the same as those in \cite{Hyunetal}. From the point of
 view of eikonal approximation, however, 
this is largely a coincidence: By adding 
 total derivatives, the structures of the E-type part could have 
 been different. 
	\item Explicit calculation in Matrix theory:\quad Various authors 
 performed explicit 1-loop calculation of the effective action 
with or without $\theta$'s in the eikonal approximation 
\cite{dkps, bfss, kraus, mcarthur, barrio}. At the off-shell level, 
some partial results were reported in \cite{TvR, Hata-Moriyama} 
and finally 
 the full 1-loop result, including all the fermionic terms, was 
obtained in  \cite{Kaz-Mura2}, which 
 agreed with all the previous results where comparisons could be made.
 Due to the different \lq frame' adopted, the result of \cite{Kaz-Mura2} 
 is superficially different from the one obtained here, but 
we have checked that after appropriate field redefinitions they agree 
completely  provided that we take 
\begin{align}
b=0, \quad  c = - \frac{15}{16}. 
\end{align}
\end{itemize}

Before we turn to the determination of the SUSY transformations, 
 we should make a remark. As an alert reader may have noticed already, 
 the set of all the relations imposed on the coefficient functions 
  forms an over-determined system. It can be checked that our solutions 
 obtained using a part of these relations do satisfy all the 
 rest of the equations, as they should. 
\subsection{D-type part of the Ward identities and 
determination of the SUSY transformation laws
}

Having determined the effective action from the E-type part of 
 the Ward identities, we now solve 
the remaining D-type part of the identities to obtain the 
form of the SUSY transformation laws. 

We start with the analysis of the part containing one power of $\theta$. 
Using the D-type terms left in  (\ref{etypep4th1}), 
 the  Ward identities of the type (\ref{dd1}) and (\ref{dd2}) at
$\calO(\p^4\theta)$ are given (in a combined form) by   
 \begin{align}
& - \Omega^{\del^2 \theta}_{m \alpha} \epsilon_\alpha a^m 
 + T^{\del^3}_{\alpha \beta} \epsilon_\beta \dot \theta_\alpha
    + 2\,i\,G^{{\p}^3{\theta}^2}\underscore1\,v^2\,{a_i}\,
     (\epsilon {{\gamma }^i}\theta ) \brkeq  +
    4\,i\,G^{{\p}^3{\theta}^2}\underscore1\,{v_i}\,
     (v \cdot a)\,(\epsilon {{\gamma }^i}\theta ) 
   - i 
\left( 4\,f^{{\p}^4}\underscore1 - 2\,G^{{\p}^3{\theta}^2}\underscore1 \right)
\,v^2\,{v_i}\,
     (\epsilon {{\gamma }^i}{\dot{\theta}}) \simeq 0 .
\end{align}
By substituting the results 
 (\ref{fp4}) and (\ref{fp3th2}) into the above equation, and reading
 off the coefficients of
$a^m$ and $\theta_\alpha$, we immediately obtain 
\begin{align}
\Omega_{m \beta}^{\p^2 \theta} \epsilon_\beta 
& = 2\,\imag \,\left( b + \frac{c}{r^7} \right) \,{v_i}\,{v_m}\,
   (\epsilon {{\gamma }^{i}}\theta ) + 
  \imag \,\left( b + \frac{c}{r^7} \right) \,v^2\,
   (\epsilon {{\gamma }^{m}}\theta ) , \label{strp2th1}
\\
T_{\alpha\beta}^{\p^3} \epsilon_\beta 
& = 3\,\imag \,\left( b + \frac{c}{r^7} \right) \,v^2\,{v_i}\,
  {{(\epsilon {{\gamma }^{i}})}_{\alpha }}  \label{stthp3th0}.
\end{align}

Likewise, using the D-type terms in (\ref{etypeth3}) and 
 the knowledge of the coefficients (\ref{fp3th2}) and  (\ref{fp2th4}), 
we get, from the Ward identity at $\calO(\p^3\theta^3)$, 
\begin{align}
\Omega_{m \beta}^{\p \theta^3} \epsilon_\beta  
=  \and \frac{7\,i\,c\,{r_i}\,{v_j}\,
       (\epsilon {{\gamma }^m}\theta )\,
       (\theta {{\gamma }^{ij}}\theta )}{2\,r^9}+
    \frac{7\,i\,c\,{r_i}\,{v_j}\,
       (\epsilon {{\gamma }^j}\theta )\,
       (\theta {{\gamma }^{im}}\theta )}{2\,r^9} , 
 \label{strp1th3}\\
T_{\alpha\beta}^{\p^2 \theta^2} \epsilon_\beta 
= \and \frac{7\,i\,c\,v^2\,{r_i}\,
       (\theta {{\gamma }^{ij}}\theta )\,
       {{(\epsilon {{\gamma }^{j}})}_{\alpha }}}{2\,r^9}+
    \frac{7\,i\,c\,{r_i}\,{v_j}\,{v_k}\,
       (\theta {{\gamma }^{ik}}\theta )\,
       {{(\epsilon {{\gamma }^{j}})}_{\alpha }}}{2\,r^9} \nn \\
& -
    \frac{7\,i\,c\,{r_i}\,{v_j}\,{v_k}\,
       (\epsilon {{\gamma }^j}\theta )\,
       {{(\theta {{\gamma }^{ik}})}_{\alpha }}}{r^9} 
 \label{stthp2th2}.
\end{align}

The procedures to get the SUSY transformation laws 
  at $\calO(\p^2\theta^5)$ and $\calO(\p \theta^7)$ are 
entirely similar. The results, which are rather involved, 
 are recorded in   Appendix B. 

\subsection{Closure relations on $\theta_\alpha$}

The final step of our endeavor is to show that the transformation 
 laws obtained above are bonafide those of supersymmetry, \ie 
 they satisfy the proper closure relations (\ref{clth}) and (\ref{clr}). 
In the course of this demonstration, we will be able to fix the 
 off-shell coefficients $A\sim D$ completely. 

In this subsection, we study the closure relation (\ref{clth}) 
 on $\theta_\alpha$.  To this end, we substitute into 
(\ref{clth}) 
the explicit expression (\ref{comth}) for the LHS,
 the expansions (\ref{effexp4}),
(\ref{expstr2}), (\ref{expstth2}) and the results obtained at order 2, 
 namely (\ref{efforder2}), (\ref{susytreer}) and  (\ref{susytreeth}). 
 Collecting 
terms with the same number of $\theta$'s, the closure relation can then be
split into the following four equations: \\ $\calO (\del^3 \theta)$:
\begin{align}
\bigg\{
i (\gamma^m \theta)_\beta 
 \frac{\delta T^{\del^3}_{\alpha \gamma}}{\delta r^m} - 
 i \vslash_{\beta \delta}
 \frac{\delta T^{\del^2 \theta^2}_{\alpha \gamma}}{\delta \theta_\delta}
& +
i \dot \Omega^{\del^2 \theta}_{m  \gamma} \gamma^m_{\alpha \beta} \bigg\}
+  ( \beta \leftrightarrow \gamma )  \nn \\
& =   
A^{0}_{\alpha \beta \gamma \delta} 
\frac{\delta \Gamma^{\del^3 \theta^2}}{\delta \theta_\delta}
+
A^{\del^2}_{\alpha \beta \gamma \delta} \dot \theta_\delta
-
B_{\alpha \beta \gamma n}^{\p \theta}  a^n 
\label{clthp3th1},
\end{align}
$\calO (\del^2 \theta^3)$:
\begin{align}
\bigg\{ 
i (\gamma^m \theta)_\beta 
 \frac{\delta T^{\del^2 \theta^2}_{\alpha \gamma}}{\delta r^m}
 - 
 i \vslash_{\beta \delta}
 \frac{\delta T^{\del \theta^4}_{\alpha \gamma}}{\delta \theta_\delta}
& +
i \dot \Omega^{\del \theta^3}_{m \gamma} \gamma^m_{\alpha \beta} 
\bigg\} +  ( \beta \leftrightarrow \gamma ) \nn \\
& =   
A^{0}_{\alpha \beta \gamma \delta} 
\frac{\delta \Gamma^{\del^2 \theta^4}}{\delta \theta_\delta}
+
A^{\del \theta^2 }_{\alpha \beta \gamma \delta} \dot \theta_\delta
-
B_{\alpha \beta \gamma n}^{\theta^3}  a^n  \label{clthp2th3}, 
\end{align}
$\calO (\del \theta^5)$:
\begin{align}
\bigg\{ i (\gamma^m \theta)_\beta 
 \frac{\delta T^{\del \theta^4}_{\alpha \gamma}}{\delta r^m}
 - 
 i \vslash_{\beta \delta}
 \frac{\delta T^{\theta^6}_{\alpha \gamma}}{\delta \theta_\delta}
+
i \dot \Omega^{\theta^5}_{m \gamma} \gamma^m_{\alpha \beta} \bigg\}
+  ( \beta \leftrightarrow \gamma ) 
& =   
A^{0}_{\alpha \beta \gamma \delta} 
\frac{\delta \Gamma^{\del \theta^6}}{\delta \theta_\delta}
+
A^{\theta^4 }_{\alpha \beta \gamma \delta} \dot \theta_\delta 
\label{clthp1th5},
\end{align}
$\calO (\theta^7)$:
\begin{align}
i (\gamma^m \theta)_\beta 
 \frac{\delta T^{\theta^6}_{\alpha \gamma}}{\delta r^m}
+  ( \beta \leftrightarrow \gamma ) 
=   
A^{0}_{\alpha \beta \gamma \delta} 
\frac{\delta \Gamma^{\theta^8}}{\delta \theta_\delta} 
\label{clthp0th7}, 
\end{align}
where $A^{0}_{\alpha \beta \gamma \delta}$ is already given in (\ref{A0}).
For later convenience we have stripped off the arbitrary spinors
 $\epsilon_\beta$ and $\lambda_\gamma$. 
 Below, we will examine the consistency of each of these 
relations  and determine the form of the remaining off-shell coefficients.

\subsubsection{Analysis 
at $\calO(\p^3 \theta)$}

First, we analyze the closure relation at $\calO(\p^3\theta)$. 
Substituting the SUSY transformation laws (\ref{strp2th1}),
(\ref{stthp3th0}) and (\ref{stthp2th2}) into the  LHS  of
(\ref{clthp3th1}) and contracting with arbitrary spinors 
$\ep_\be \lam_\ga \psi_\al$ from left,  we get 
\begin{align}
\text{LHS of 
(\ref{clthp3th1})}=   \and -\frac{7\,c\,v^2\,{r_i}\,{v_j}\,
       (\epsilon {{\gamma }^j}\theta )\,
       (\lambda {{\gamma }^i}\psi )}{r^9}-
    \frac{7\,c\,v^2\,{r_i}\,{v_j}\,
       (\epsilon {{\gamma }^j}\psi )\,
       (\lambda {{\gamma }^i}\theta )}{r^9}+
    \frac{7\,c\,v^2\,{r_i}\,{v_j}\,
       (\epsilon {{\gamma }^i}\theta )\,
       (\lambda {{\gamma }^j}\psi )}{r^9} \brkeq 
   +\frac{7\,c\,v^2\,{r_i}\,{v_j}\,
       (\epsilon {{\gamma }^i}\psi )\,
       (\lambda {{\gamma }^j}\theta )}{r^9}-
    \frac{7\,c\,v^2\,{r_i}\,{v_j}\,
       (\epsilon {{\gamma }^{ijk}}\theta )\,
       (\lambda {{\gamma }^k}\psi )}{r^9} 
\brkeq  +
    \frac{7\,c\,v^2\,{r_i}\,{v_j}\,
       (\epsilon {{\gamma }^k}\psi )\,
       (\lambda {{\gamma }^{ijk}}\theta )}{r^9} 
   +\frac{14\,c\,v^2\,{r_i}\,{v_j}\,(\epsilon \lambda )\,
       (\psi {{\gamma }^{ij}}\theta )}{r^9} \brkeq 
+ 2\,\left( b + \frac{c}{r^7} \right) \,(v \cdot a)\,
   (\epsilon {{\gamma }^{{a_1}}}\theta )\,
   (\lambda {{\gamma }^{{a_1}}}\psi ) - 
  2\,\left( b + \frac{c}{r^7} \right) \,v^2\,
   (\epsilon {{\gamma }^{{a_1}}}{\dot{\theta}})\,
   (\lambda {{\gamma }^{{a_1}}}\psi ) \brkeq 
- 
  2\,\left( b + \frac{c}{r^7} \right) \,(v \cdot a)\,
   (\epsilon {{\gamma }^{{a_1}}}\psi )\,
   (\lambda {{\gamma }^{{a_1}}}\theta ) + 
  2\,\left( b + \frac{c}{r^7} \right) \,v^2\,
   (\epsilon {{\gamma }^{{a_1}}}\psi )\,
   (\lambda {{\gamma }^{{a_1}}}{\dot{\theta}}) \brkeq  + 
  2\,\left( b + \frac{c}{r^7} \right) \,{v_{{i_1}}}\,
   {a_{{i_2}}}\,(\epsilon {{\gamma }^{{i_2}}}\theta )\,
   (\lambda {{\gamma }^{{i_1}}}\psi ) - 
  2\,\left( b + \frac{c}{r^7} \right) \,{v_{{i_1}}}\,
   {a_{{i_2}}}\,(\epsilon {{\gamma }^{{i_2}}}\psi )\,
   (\lambda {{\gamma }^{{i_1}}}\theta ) \brkeq + 
  2\,\left( b + \frac{c}{r^7} \right) \,{v_{{i_1}}}\,
   {a_{{i_2}}}\,(\epsilon {{\gamma }^{{i_1}}}\theta )\,
   (\lambda {{\gamma }^{{i_2}}}\psi ) - 
  4\,\left( b + \frac{c}{r^7} \right) \,{v_{{i_1}}}\,
   {v_{{i_2}}}\,(\epsilon {{\gamma }^{{i_1}}}{\dot{\theta}}
    )\,(\lambda {{\gamma }^{{i_2}}}\psi ) \brkeq  - 
  2\,\left( b + \frac{c}{r^7} \right) \,{v_{{i_1}}}\,
   {a_{{i_2}}}\,(\epsilon {{\gamma }^{{i_1}}}\psi )\,
   (\lambda {{\gamma }^{{i_2}}}\theta ) + 
  4\,\left( b + \frac{c}{r^7} \right) \,{v_{{i_1}}}\,
   {v_{{i_2}}}\,(\epsilon {{\gamma }^{{i_1}}}\psi )\,
   (\lambda {{\gamma }^{{i_2}}}{\dot{\theta}}) \period
\label{lhsclth1}
\end{align}
 Note that the first 7 terms are of E-type and the rest are 
 of D-type. Turning to the RHS of (\ref{clthp3th1}), 
the first term can be easily computed  using the explicit
 expression $\Gamma^{\p^3\theta^2} = \int \! \d\tau \,
7\,c\,v^2\,{r_i}\,{v_j}\,
    (\theta {{\gamma }^{ij}}\theta )/(2\,r^9)$. The result is
\begin{align}
\psi_\alpha \epsilon_\beta \lambda_\gamma 
A^{0}_{\alpha\beta\gamma\delta}
\frac{\delta \Gamma^{\p^3\theta^2}}{\delta \theta^\delta} 
=   \and -\frac{7\,c\,v^2\,{r_i}\,{v_j}\,(\lambda \psi )\,
       (\epsilon {{\gamma }^{ij}}\theta )}{r^9}+
    \frac{7\,c\,v^2\,{r_i}\,{v_j}\,(\epsilon \psi )\,
       (\lambda {{\gamma }^{ij}}\theta )}{r^9} \brkeq  
-
    \frac{7\,c\,v^2\,{r_i}\,{v_j}\,
       (\epsilon {{\gamma }^j}\lambda )\,
       (\psi {{\gamma }^i}\theta )}{r^9} 
   +\frac{7\,c\,v^2\,{r_i}\,{v_j}\,
       (\epsilon {{\gamma }^i}\lambda )\,
       (\psi {{\gamma }^j}\theta )}{r^9} \brkeq  +
    \frac{7\,c\,v^2\,{r_i}\,{v_j}\,(\epsilon \lambda )\,
       (\psi {{\gamma }^{ij}}\theta )}{r^9}+
    \frac{7\,c\,v^2\,{r_i}\,{v_j}\,
       (\epsilon {{\gamma }^k}\lambda )\,
       (\psi {{\gamma }^{ijk}}\theta )}{r^9} 
\label{rhsfclth1} \period
\end{align}
Since the remaining two terms on the RHS of (\ref{clthp3th1}) are
both of D-type,  E-type terms in (\ref{lhsclth1}) and
(\ref{rhsfclth1}) must cancel for consistency.
Though it is not self-evident, we can show, 
with the help of Fierz identities,  that they do cancel each other. 

We are thus left with purely D-type terms on both sides, and 
we can easily read off $A^{\p^2}_{\alpha\beta\gamma\delta}$ and $B^{\p
\theta}_{\alpha\beta\gamma m}$ from this relation:
\begin{align}
 A^{\p^2}_{\alpha\beta\gamma\delta} =  & 
2\,\left( b + \frac{c}{r^7} \right) \,v^2\,
   {{{{\gamma }^{i}}}_{\beta \delta }}\,
   {{{{\gamma }^{i}}}_{\gamma \alpha }} - 
  2\,\left( b + \frac{c}{r^7} \right) \,v^2\,
   {{{{\gamma }^{i}}}_{\beta \alpha }}\,
   {{{{\gamma }^{i}}}_{\gamma \delta }} \brkeq + 
  4\,\left( b + \frac{c}{r^7} \right) \,{v_i}\,{v_j}\,
   {{{{\gamma }^{i}}}_{\beta \delta }}\,
   {{{{\gamma }^{j}}}_{\gamma \alpha }} - 
  4\,\left( b + \frac{c}{r^7} \right) \,{v_i}\,{v_j}\,
   {{{{\gamma }^{i}}}_{\beta \alpha }}\,
   {{{{\gamma }^{j}}}_{\gamma \delta }} ,
\end{align}
\begin{align}
B^{\p \theta}_{\alpha\beta\gamma m} 
= & 2\,\left( b + \frac{c}{r^7} \right) \,{v_m}\,
   {{{{\gamma }^{i}}}_{\alpha \beta }}\,
   {{(\theta {{\gamma }^{i}})}_{\gamma }} - 
  2\,\left( b + \frac{c}{r^7} \right) \,{v_m}\,
   {{{{\gamma }^{i}}}_{\alpha \gamma }}\,
   {{(\theta {{\gamma }^{i}})}_{\beta }} \brkeq  - 
  2\,\left( b + \frac{c}{r^7} \right) \,{v_i}\,
   {{{{\gamma }^{i}}}_{\alpha \gamma }}\,
   {{(\theta {{\gamma }^{m}})}_{\beta }} 
+ 
  2\,\left( b + \frac{c}{r^7} \right) \,{v_i}\,
   {{{{\gamma }^{i}}}_{\alpha \beta }}\,
   {{(\theta {{\gamma }^{m}})}_{\gamma }}  \brkeq  + 
  2\,\left( b + \frac{c}{r^7} \right) \,{v_i}\,
   {{{{\gamma }^{m}}}_{\alpha \beta }}\,
   {{(\theta {{\gamma }^{i}})}_{\gamma }} - 
  2\,\left( b + \frac{c}{r^7} \right) \,{v_i}\,
   {{{{\gamma }^{m}}}_{\alpha \gamma }}\,
   {{(\theta {{\gamma }^{i}})}_{\beta }} .
\end{align}
This completes the analysis at $\calO(\p^3 \theta)$. 

\subsubsection{Analysis at $\calO(\p^2 \theta^3)$}

Although the procedure is entirely similar as above, the amount of 
 computations  needed 
 at $\calO(\p^2 \theta^3)$ increases considerably. For example, 
 the number of E-type terms in the relation (\ref{clthp2th3})
 is 269 on the LHS and 12 on the RHS and we must show that they precisely 
cancel. By using the explicit representation of $SO(9)$ $\ga$-matrices
and with the aid of Mathematica,  we have checked that they indeed cancel. 
Once E-type terms are cancelled, determination of the form of $A^{\p
\theta^2}_{\alpha\beta\gamma\delta}$ and
$B^{\theta^3}_{\alpha\beta\gamma m}$ from the remaining D-type terms
 is not difficult. The results are
\begin{align}
& A^{\p \theta^2}_{\alpha\beta\gamma\delta} = \brkeq
-\frac{7\,c\,{r_i}\,{v_j}\,
       (\theta {{\gamma }^{ik}}\theta )\,
       {{{{\gamma }^{j}}}_{\delta \gamma }}\,
       {{{{\gamma }^{k}}}_{\alpha \beta }}}{2\,r^9}+
    \frac{7\,c\,{r_i}\,{v_j}\,
       (\theta {{\gamma }^{ik}}\theta )\,
       {{{{\gamma }^{j}}}_{\beta \delta }}\,
       {{{{\gamma }^{k}}}_{\alpha \gamma }}}{2\,r^9}+
    \frac{7\,c\,{r_i}\,{v_j}\,
       (\theta {{\gamma }^{ik}}\theta )\,
       {{{{\gamma }^{j}}}_{\alpha \gamma }}\,
       {{{{\gamma }^{k}}}_{\beta \delta }}}{2\,r^9} \brkeq 
   -\frac{7\,c\,{r_i}\,{v_j}\,
       (\theta {{\gamma }^{ik}}\theta )\,
       {{{{\gamma }^{j}}}_{\alpha \beta }}\,
       {{{{\gamma }^{k}}}_{\delta \gamma }}}{2\,r^9}+
    \frac{7\,c\,{r_i}\,{v_j}\,
       {{{{\gamma }^{k}}}_{\beta \delta }}\,
       {{(\theta {{\gamma }^{k}})}_{\gamma }}\,
       {{(\theta {{\gamma }^{ij}})}_{\alpha }}}{r^9}-
    \frac{7\,c\,{r_i}\,{v_j}\,
       {{{{\gamma }^{k}}}_{\delta \gamma }}\,
       {{(\theta {{\gamma }^{k}})}_{\beta }}\,
       {{(\theta {{\gamma }^{ij}})}_{\alpha }}}{r^9} \brkeq 
   -\frac{7\,c\,{r_i}\,{v_j}\,
       {{{{\gamma }^{k}}}_{\alpha \beta }}\,
       {{(\theta {{\gamma }^{k}})}_{\gamma }}\,
       {{(\theta {{\gamma }^{ij}})}_{\delta }}}{r^9}+
    \frac{7\,c\,{r_i}\,{v_j}\,
       {{{{\gamma }^{k}}}_{\alpha \gamma }}\,
       {{(\theta {{\gamma }^{k}})}_{\beta }}\,
       {{(\theta {{\gamma }^{ij}})}_{\delta }}}{r^9}+
    \frac{7\,c\,{r_i}\,{v_j}\,
       {{{{\gamma }^{k}}}_{\beta \delta }}\,
       {{(\theta {{\gamma }^{j}})}_{\gamma }}\,
       {{(\theta {{\gamma }^{ik}})}_{\alpha }}}{r^9} \brkeq 
   -\frac{7\,c\,{r_i}\,{v_j}\,
       {{{{\gamma }^{k}}}_{\delta \gamma }}\,
       {{(\theta {{\gamma }^{j}})}_{\beta }}\,
       {{(\theta {{\gamma }^{ik}})}_{\alpha }}}{r^9}-
    \frac{7\,c\,{r_i}\,{v_j}\,
       {{{{\gamma }^{k}}}_{\alpha \beta }}\,
       {{(\theta {{\gamma }^{j}})}_{\gamma }}\,
       {{(\theta {{\gamma }^{ik}})}_{\delta }}}{r^9}+
    \frac{7\,c\,{r_i}\,{v_j}\,
       {{{{\gamma }^{k}}}_{\alpha \gamma }}\,
       {{(\theta {{\gamma }^{j}})}_{\beta }}\,
       {{(\theta {{\gamma }^{ik}})}_{\delta }}}{r^9} ,
\end{align}
\begin{align}
B^{\theta^3}_{\alpha\beta\gamma m} 
=  \and +\frac{7\,c\,{r_i}\,(\theta {{\gamma }^{im}}\theta )\,
       {{{{\gamma }^{j}}}_{\alpha \beta }}\,
       {{(\theta {{\gamma }^{j}})}_{\gamma }}}{2\,r^9}-
    \frac{7\,c\,{r_i}\,(\theta {{\gamma }^{im}}\theta )\,
       {{{{\gamma }^{j}}}_{\alpha \gamma }}\,
       {{(\theta {{\gamma }^{j}})}_{\beta }}}{2\,r^9} \brkeq 
   -\frac{7\,c\,{r_i}\,(\theta {{\gamma }^{ij}}\theta )\,
       {{{{\gamma }^{j}}}_{\alpha \gamma }}\,
       {{(\theta {{\gamma }^{m}})}_{\beta }}}{2\,r^9}+
    \frac{7\,c\,{r_i}\,(\theta {{\gamma }^{ij}}\theta )\,
       {{{{\gamma }^{j}}}_{\alpha \beta }}\,
       {{(\theta {{\gamma }^{m}})}_{\gamma }}}{2\,r^9} .
\end{align}

\subsubsection{Analysis at $\calO(\p \theta^5)$}

The situation is quite analogous to the one just described, except 
 that it is even more involved. 
We have found that 393 and 35 E-type terms on the LHS and RHS,
 respectively,  of the relation 
(\ref{clthp1th5}) cancel exactly and the relation among the remaining 
 D-type terms fixes the form of $A^{\theta^4}_{\alpha\beta\gamma\delta}$
uniquely. Unfortunately, the result consists of  152
terms, which is too space-consuming to be displayed in this paper. 

\subsubsection{Analysis at $\calO(\theta^7)$}

The final closure relation (\ref{clthp1th5}) left to be examined 
consists only of E-type terms and 
does not contain any of the 
 off-shell coefficients. Thus it serves as  a consistency check of our 
SUSY transformation laws.  
The LHS of (\ref{clthp1th5}) has 85 E-type terms while its RHS has 
35. In a manner similar to the previous analyses, we have checked that 
these E-type terms match precisely. 

This completes the analysis 
of the closure relation on $\theta_\al$. 

\subsection{Closure relations on $r^m$}

To finish up our rather long exploration, we examine
 the closure relation (\ref{clr}) on $r^m$.

By using the explicit expression (\ref{comr}), the expansions 
(\ref{effexp4}), (\ref{expstr2}), (\ref{expstth2}) and the results 
(\ref{efforder2}), (\ref{susytreer}), (\ref{susytreeth}) obtained at
order 2, 
we can decompose the closure relation (\ref{clr}) into 
the following 4 equations:\parmedskipn
$\calO (\del^3)$:
\begin{align}
\bigg\{ i \vslash_{\alpha \gamma } 
 \frac{\delta \Omega_{m \beta}^{\del^2 \theta}}{\delta \theta_\delta}
+ 
i \gamma^m_{\alpha\gamma} T^{\del^3}_{\gamma \beta} \bigg\}
+ ( \alpha \leftrightarrow \beta  )  
  = - D^\del_{m \alpha \beta n}  a^n  , 
\label{clrp3}
\end{align}
$\calO (\del^2 \theta^2)$:
\begin{align}
\bigg\{ i (\gamma^n \theta)_\alpha 
 \frac{\delta \Omega^{\del^2 \theta}_{m \beta }}{\delta r^n}
+
 i \vslash_{\alpha \gamma } 
 \frac{\delta \Omega_{m \beta}^{\del \theta^3}}{\delta \theta_\gamma}
+ 
i \gamma^m_{\alpha\gamma} T^{\del^2 \theta^2}_{\gamma \beta} \bigg\}
+ ( \alpha \leftrightarrow \beta  )  
  = C_{m \alpha \beta \delta}^{\p\theta}  \dot \theta_\delta 
     - D^{\theta^2}_{m \alpha \beta n}  a^n  , 
\label{clrp2}
\end{align}
$\calO (\del \theta^4)$:
\begin{align}
\bigg\{ i (\gamma^n \theta)_\alpha 
 \frac{\delta \Omega^{\del \theta^3}_{m \beta }}{\delta r^n}
+
 i \vslash_{\alpha \gamma } 
 \frac{\delta \Omega_{m \beta}^{\theta^5}}{\delta \theta_\gamma}
+ 
i \gamma^m_{\alpha\gamma} T^{\del \theta^4}_{\gamma \beta} \bigg\}
+ ( \alpha \leftrightarrow \beta  )  
& = C_{m \alpha \beta \delta}^{\theta^3}  \dot \theta_\delta ,
\label{clrp1}
\end{align}
$\calO (\theta^6)$:
\begin{align}
\bigg\{ i (\gamma^n \theta)_\alpha 
 \frac{\delta \Omega^{\theta^5}_{m \beta }}{\delta r^n}
+ 
i \gamma^m_{\alpha\gamma} T^{\theta^6}_{\gamma \beta} \bigg\}
+ ( \alpha \leftrightarrow \beta  )  
& = 0  \period
\label{clrp0}
\end{align}

At $\calO(\p^3)$, 
substituting 
the SUSY transformation laws (\ref{strp2th1}), (\ref{stthp3th0})
into (\ref{clrp3}), one can easily find 
\begin{align}
 D^{\p}_{m \alpha\beta n} = 0 .
\end{align}

Similarly, at $\calO(\p^2 \theta^2)$, substituting (\ref{strp2th1}),
(\ref{strp1th3}), (\ref{stthp2th2}) into (\ref{clrp2}), we obtain
\begin{align}
D^{\theta^2}_{m \alpha\beta n} = 0\comma 
\end{align}
and 
\begin{align}
 C^{\p \theta}_{m\beta\gamma\delta} = \and 
-2\,\left( b + \frac{c}{r^7} \right) \,{v_m}\,
   {{{{\gamma }^{i}}}_{\beta \delta }}\,
   {{(\theta {{\gamma }^{i}})}_{\gamma }} + 
  2\,\left( b + \frac{c}{r^7} \right) \,{v_m}\,
   {{{{\gamma }^{i}}}_{\delta \gamma }}\,
   {{(\theta {{\gamma }^{i}})}_{\beta }} \brkeq  + 
  2\,\left( b + \frac{c}{r^7} \right) \,{v_i}\,
   {{{{\gamma }^{i}}}_{\delta \gamma }}\,
   {{(\theta {{\gamma }^{m}})}_{\beta }} 
-   2\,\left( b + \frac{c}{r^7} \right) \,{v_i}\,
   {{{{\gamma }^{i}}}_{\beta \delta }}\,
   {{(\theta {{\gamma }^{m}})}_{\gamma }} \brkeq  - 
  2\,\left( b + \frac{c}{r^7} \right) \,{v_i}\,
   {{{{\gamma }^{m}}}_{\beta \delta }}\,
   {{(\theta {{\gamma }^{i}})}_{\gamma }} + 
  2\,\left( b + \frac{c}{r^7} \right) \,{v_i}\,
   {{{{\gamma }^{m}}}_{\delta \gamma }}\,
   {{(\theta {{\gamma }^{i}})}_{\beta }} .
\end{align}

Beginning at $\calO(\p \theta^4)$, we need various Fierz identities.
 Substituting (\ref{strp1th3}), (\ref{strp0th5}) and (\ref{stthp1th4})
 into (\ref{clrp1}), we get 175 E-type terms and 4 D-type terms on the
 LHS, while we do not have any E-type terms on the RHS.  Thus the E-type
 terms on the LHS should vanish by themselves. As before, with the help
 of Mathematica , we can show that they indeed do. From the relations
 among the remaining D-type terms, we read off the
 $C^{\theta^3}_{m\alpha\beta\gamma}$ as
\begin{align}
C^{\theta^3}_{m\beta\gamma\delta}
 =  \and -\frac{7\,c\,{r_i}\,(\theta {{\gamma }^{im}}\theta )\,
       {{{{\gamma }^{j}}}_{\beta \delta }}\,
       {{(\theta {{\gamma }^{j}})}_{\gamma }}}{2\,r^9}+
    \frac{7\,c\,{r_i}\,(\theta {{\gamma }^{im}}\theta )\,
       {{{{\gamma }^{j}}}_{\delta \gamma }}\,
       {{(\theta {{\gamma }^{j}})}_{\beta }}}{2\,r^9} \brkeq 
   +\frac{7\,c\,{r_i}\,(\theta {{\gamma }^{ij}}\theta )\,
       {{{{\gamma }^{j}}}_{\delta \gamma }}\,
       {{(\theta {{\gamma }^{m}})}_{\beta }}}{2\,r^9}-
    \frac{7\,c\,{r_i}\,(\theta {{\gamma }^{ij}}\theta )\,
       {{{{\gamma }^{j}}}_{\beta \delta }}\,
       {{(\theta {{\gamma }^{m}})}_{\gamma }}}{2\,r^9} .
\end{align}

Finally at $\calO(\theta^6)$, the relevant closure relation (\ref{clrp0}) 
 does not contain any off-shell coefficients and hence it only provides 
 a consistency check.  Calculating the LHS of (\ref{clrp0}) using 
(\ref{strp0th5}) and (\ref{stthp0th6}), we get 97 E-type terms.  It can be
 shown that these terms cancel out due to Fierz
identities.

\section{Summary and Discussions}

In this paper we have developed an efficient unambiguous scheme 
to analyze the SUSY Ward identity for the effective action,
 the SUSY transformations  and  their closure relations 
to clarify the role of maximal supersymmetry in the dynamics 
 of a D-particle. Our analysis is valid for 
 completely off-shell configurations and assumes no knowledge of
 the underlying theory. 

We found that the effective actions
 at order 2 and at order 4 are completely determined, up to 
 two numerical constants, by the symmetry requirements alone. 
In the context of Matrix theory for M theory, this provides 
 a complete unambiguous proof of off-shell non-renormalization 
 theorems. 

Moreover, in contrast to previous investigations, 
we have been able to determine the SUSY transformations 
 uniquely and proved that they satisfy the proper closure relations. 
This includes the determination of the off-shell coefficient functions
 appearing in the closure relation as well. As far as the 
 system under consideration is concerned, we believe that our analysis 
 has fully elucidated the power of the symmetries, in particular 
 the supersymmetry.

A natural extension of this work would be the generalization to higher
orders in the derivative expansion. For example, let us consider the
effective action at order 6. The purely bosonic part at 1-loop 
 was computed in \cite{okawa9903} in Matrix theory and a crude analysis 
without assuming such an underlying theory has been attempted in 
 \cite{Pabanetal2}. To perform a complete analysis, we need to 
examine the relevant Ward identity, which schematically is of the form 
\begin{align}
 \delta^{(0)}_\epsilon \Gamma^{(6)} + 
 \delta^{(2)}_\epsilon  \Gamma^{(4)} +
 \delta^{(4)}_\epsilon \Gamma^{(2)} = 0 ,
\label{ward6}
\end{align}
where $\Gamma^{(6)}$ is the effective action at order 6 and
$\delta^{(4)}_\epsilon$ is the SUSY transformation at order 4.
  Since we now have the explicit form 
 of $\delta^{(2)}_\ep$ and $\Gamma^{(4)}$, the second term on the LHS can 
 be computed. Further, similarly to the case of $\Ga^{(4)}$, 
  $\Gamma^{(6)}$ can be brought to the form 
\begin{align}
 \Gamma^{(6)} = \int \!\d\tau \left( 
\bar \calL^{(6)} + 
a_m X_m^{(4)} - \Psi_\alpha^{(9/2)} \dot \theta_\alpha
\right) , 
\end{align}
where $\bar \calL^{(6)}$ denotes the purely E-type part. This means 
 that, by the use of E-type -  D-type separation method developed in 
 this paper, it should be possible to determine $\Ga^{(6)}$ and 
 $\delta^{(4)}_\ep$. Moreover, since $\delta^{(2)}_\epsilon  \Gamma^{(4)}$
part acts as an \lq inhomogeneous term' in the relevant equations, 
even the normalization of $\Ga^{(6)}$ is expected to be fixed 
by that of $\Ga^{(4)}$. This is a new situation starting at this order. 
The actual calculation would require considerable effort, however. 

Another important direction into which to extend our work is to apply
our scheme to the multi-body system.  Although performed in the 
 eikonal approximation, an explicit calculation in Matrix theory 
 revealed \cite{Okawa-Yoneya, Okawa-Yoneya2} 
that even the non-linear part of the 11-dimensional 
supergravity interactions are correctly encoded in Matrix theory. 
It is extremely important to clarify to what extent this feature 
 is due to supersymmetry. Again, practically this requires a vast amount 
of work mainly because the number of possible terms in various quantities 
 increases significantly compared to the two-body case. 

We hope that progress on these issues can be made in future investigations.

\bigskip
\bigskip
\bigskip

\par\bigskip\noindent
{\large\bf Acknowledgment}\par\smallskip\noindent
We wish to acknowledge valuable discussions with Y. Okawa.
The research of Y.K. is supported in part by the 
 Grant-in-Aid for Scientific Research (B) 
No.~12440060, while that of T.M. is supported in part by 
the Japan Society for Promotion of Science under the Predoctoral
Research Program No.~12-9617, 
both from the Japan  Ministry of Education,  Science
 and Culture. 
\newpage

\appendix

\setcounter{equation}{0}
\renewcommand{\theequation}{A.\arabic{equation}}
\section*{Appendix A: \ \
Null transformations and their closure relations}

In this appendix, we study the null transformations 
$\Delta_\epsilon^{(2)}$, which are the solutions of the equation
\begin{align}
-a_m \Dept r_m +(\Dept \theta_\al) \thdot_\al =\frac{\d G(\tau)}{\d \tau} ,
\label{nulleq}
\end{align}
for some $G$, and clarify how they affect the closure relations.

\bigskip
\noindent
{\bf Enumeration of null transformations}

An efficient algorithm for finding solutions to (\ref{nulleq}) 
is to write down the most general
 form of $\Dept \theta_\al$ and  see if $(\Dept \theta_\al)
 \thdot_\al$ can be rewritten completely into the form $a_m X_m$ by
 integration by parts.  When that is possible, we get a solution 
 by setting $\Dept r_m = X_m$.  

We now enumerate all possible solutions. 
Since $\Dept \theta_\al$ is of order 3, apart from $\ep_\al$, it may 
 contain derivatives of $r_m$  up to $\dot{a}_m$. 

\begin{enumerate}
\item   First, consider 
the case $\Dept \theta_\al = \dot{a}_m X_{m\al}$.  By integration by parts, 
we can rewrite $\dot{a}_m X_{m\al}\thdot_\al$ into $-a_m
\del_\tau(X_{m\al}\thdot_\al) $. So there is always a solution.
\item Next consider the case where $\Dept \theta_\al = a_m
Y_{m\al}$. Then $\Dept r_m = Y_{m\al}\thdot_\al$  always gives
a solution (with $G=0$).  
\item The remaining case is the one in which 
 $\Dept \theta_\al$ does not contain
$a_m$. There are two possibilities: \parsmallskipn
(1) \ One possibility of rewriting $(\Dept \theta_\al) \thdot_\al$
{\it entirely} into the form $a_m X_m$ occurs when $\Dept \theta_\al$ consists
 of $v_m$ only,  since then integration by parts always produces a factor 
 of $a_m$.  Since the order of $\Dept \theta_\al$ is 3, we must use three 
$v_m$'s and $\ep_\al$. The only possibility is $\Dept \theta_\al = k_1 v^2
(\slash{v}\ep)_\al$ where $k_1$ is a numerical constant. Then, by
	performing
 integration by parts, we find 
\begin{align}
\Dept r_m = -2 k_1 v_m (\slash{v}\ep)_\al\theta_\al
 -kv^2 (\ga_m\ep)_\al
\theta_\al .
\end{align}
(2) \ The second possibility is when $\Delta^{(2)}_\epsilon r^m$ 
consists of $\theta$ only, since after integrating 
$a^m \Delta^{(2)}_\epsilon r^m$ by parts the terms
      containing $\dot \theta$
may be canceled by $(\Delta_\epsilon^{(2)} \theta_\alpha) \dot \theta_\alpha$.
Taking into account its order and C-symmetry  requirement, 
the only possibility is 
\begin{align}
 \Dept r_m &= k_2 (\epsilon {{\gamma }^n}\theta )\,
    (\theta {{\gamma }^{am}}\theta )\,
    (\theta {{\gamma }^{an}}\theta ) ,
\end{align}
where $k_2$ is a numerical constant. Performing integration by parts, we
      find that the following $\Dept \theta_\al$ gives a solution:
\begin{align}
\Dept \theta_\al &= 
- k_2 (\epsilon {{\gamma }^n})_\alpha\,
     (\theta {{\gamma }^{am}}\theta )\,
     (\theta {{\gamma }^{an}}\theta )\,{v_m} - 
  2\,k_2\,(\epsilon {{\gamma }^n}\theta )\,
   (\theta {{\gamma }^{an}}\theta )\,
   (\theta {{\gamma }^{am}})_\alpha\,{v_m} \nn \\ & \quad  - 
  2\,k_2\,(\epsilon {{\gamma }^n}\theta )\,
   (\theta {{\gamma }^{am}}\theta )\,
   (\theta {{\gamma }^{an}})_\alpha \,{v_m} .  
\end{align}
\end{enumerate}

Summarizing, there are 4 types of solutions:
\begin{align}
{\rm (i)}\qquad \Dept \theta_\al &= \dot{a}_m X_{m\al\be}\ep_\be ,\\
\Dept r_m &= -\del_\tau (X_{m\al\be}\ep_\be \thdot_\al),\\
G &= a_m X_{m\al\be}\ep_\be \thdot_\al ,\\
{\rm (ii)}\qquad \Dept \theta_\al &= a_m Y_{m\al\be}\ep_\be , \\
\Dept r_m &= Y_{m\al\be}\ep_\be \thdot_\al ,\\
G &= 0 , \\
{\rm (iii)} \qquad \Dept \theta_\al &= k_1 v^2 (\slash{v}\ep)_\al ,
\label{case3r}
\\
\Dept r_m &= -2k_1 v_m \ep \slash{v}\theta
 -k_1 v^2 \ep \ga_m \theta , \label{case3th}\\
G &= k_1 v^2 \ep \slash{v} \theta ,\\
{\rm (iv)}\qquad 
\Dept \theta_\al &= 
- k_2 (\epsilon {{\gamma }^n})_\alpha\,
     (\theta {{\gamma }^{am}}\theta )\,
     (\theta {{\gamma }^{an}}\theta )\,{v_m} - 
  2\, k_2 \,(\epsilon {{\gamma }^n}\theta )\,
   (\theta {{\gamma }^{an}}\theta )\,
   (\theta {{\gamma }^{am}})_\alpha\,{v_m} \nn \\ & \quad  - 
  2\, k_2 \,(\epsilon {{\gamma }^n}\theta )\,
   (\theta {{\gamma }^{am}}\theta )\,
   (\theta {{\gamma }^{an}})_\alpha \,{v_m},
\label{case4r} \\
\Dept r_m &= k_2 (\epsilon {{\gamma }^n}\theta )\,
    (\theta {{\gamma }^{am}}\theta )\,
    (\theta {{\gamma }^{an}}\theta ) ,
 \label{case4th}\\
G &= k_2 (\epsilon {{\gamma }^n}\theta )\,
  (\theta {{\gamma }^{am}}\theta )\,
  (\theta {{\gamma }^{an}}\theta )\,{v_m} .
\end{align}

\bigskip
\noindent
{\bf Examination of the closure relation}

Now we study how these null transformations affect the closure relations. 
Let $\delta_\ep = \depz + \dept$ be a SUSY transformation which 
already satisfies the proper closure relations. Then, an addition of 
 $\Dept$ produces, at order 2, a contribution to the commutator 
\begin{eqnarray}
\left( \left[ \depz, \Dlamt\right] -(\ep \leftrightarrow \lam)
\right)
\vecii{r_m}{\theta_\al}  . 
\end{eqnarray}
We shall examine if this is of an appropriate form for proper closure 
 relations to be maintained, for each of the 4 solutions above. 
\renewcommand{\theenumi}{\roman{enumi}}
\renewcommand{\labelenumi}{(\theenumi)}

\begin{enumerate}
 \item \quad $X_{m\alpha\beta}$ is an arbitrary structure of
order 0. Combined with  the restriction from C-symmetry, the only
       possible structure for $X_{m\alpha\beta}$ is
\begin{align}
X_{m \alpha\beta} = {X_1}\,{{\gamma }^{m}_{\alpha \beta }} ,
\end{align}
where  ${X_1}$ is a function of $r(\tau)$ only.
Then, the SUSY transformation laws become
\begin{align}
\Dept r^m &=  - \frac{({r \cdot v})\,
        (\epsilon {{\gamma }^{m}}{\dot{\theta}})}{r} 
     \frac{\d {X_1}}{\d r} - 
  {X_1}\,(\epsilon {{\gamma }^{m}}{\ddot{\theta}})  , \\ 
\Dept \theta_\alpha &=  {X_1}\,{{({{\gamma }^{{i}}}\epsilon )}_{\alpha }}\,
  {{{\dot{a}}}_{{i}}} 
\end{align}
and  the additional contribution to 
the closure relation for $r^m$ takes the form
\begin{align}
\left( \left[ \depz, \Dlamt\right] -(\ep \leftrightarrow \lam)
\right)r^m =
 \frac{2\,\imag \,(\epsilon \lambda )\,({r \cdot v})\,{a_m}}
    {r}\frac{\d {X_1}}{\d r} + 
  4\,\imag \,{X_1}\,(\epsilon \lambda )\,{{{\dot{a}}}_m} .
\end{align}
While the first term, proportional to $a^m$, only modifies the form 
 of  the off-shell coefficient $D_{m \alpha \beta n}$, 
 the second term containing $\dot a^m$ cannot be absorbed into 
any of the coefficient functions and hence spoils the proper closure relation. 
Thus, the solution (i) does not qualify as proper SUSY transformation laws.
\item \quad 
 $Y_{m\alpha\beta}$ is an arbitrary structure of 
order 1. Combined with  the restriction from C-symmetry, the possible
       structures for $Y_{m\alpha\beta}$ are
\begin{align}
Y_{m \alpha\beta} =  {Y_1}\,({r \cdot v})\,{{{{\gamma }^{m}}}_{\alpha \beta }} +
  {Y_2}\,{{{{\gamma }^{mnl}}}_{\alpha \beta }}\,{r_l}\,{v_n}
+ Y_{m \alpha\beta\rho\sigma}^{\theta^2} \theta_\rho \theta_\sigma  .
\end{align}
Here $Y_{i}$, $Y_{m \alpha\beta\rho\sigma}^{\theta^2}$ are functions of
       $r(\tau)$ only and the most general form of $Y_{m
       \alpha\beta\rho\sigma}^{\theta^2}$ is
\begin{align}
Y_{m \alpha\beta\rho\sigma}^{\theta^2} = & 
Y^{{\theta}^2}\underscore1\,
   {{{{\gamma }^{i}}}_{\alpha \beta }}\,
   {{{{\gamma }^{mi}}}_{\rho \sigma }} + 
  Y^{{\theta}^2}\underscore2\,
   {{{{\gamma }^{ij}}}_{\alpha \beta }}\,
   {{{{\gamma }^{mij}}}_{\rho \sigma }} + 
  Y^{{\theta}^2}\underscore3\,
   {{{{\gamma }^{ij}}}_{\rho \sigma }}\,
   {{{{\gamma }^{mij}}}_{\alpha \beta }} + 
  Y^{{\theta}^2}\underscore4\,
   {{{{\gamma }^{ijk}}}_{\rho \sigma }}\,
   {{{{\gamma }^{mijk}}}_{\alpha \beta }} \brkeq + 
  Y^{{\theta}^2}\underscore5\,{r_j}\,{r_m}\,
   {{{{\gamma }^{i}}}_{\alpha \beta }}\,
   {{{{\gamma }^{ij}}}_{\rho \sigma }} + 
  Y^{{\theta}^2}\underscore6\,{r_i}\,{r_j}\,
   {{{{\gamma }^{i}}}_{\alpha \beta }}\,
   {{{{\gamma }^{mj}}}_{\rho \sigma }} + 
  Y^{{\theta}^2}\underscore7\,{r_k}\,{r_m}\,
   {{{{\gamma }^{ij}}}_{\alpha \beta }}\,
   {{{{\gamma }^{ijk}}}_{\rho \sigma }} \brkeq  + 
  Y^{{\theta}^2}\underscore8\,{r_k}\,{r_m}\,
   {{{{\gamma }^{ij}}}_{\rho \sigma }}\,
   {{{{\gamma }^{ijk}}}_{\alpha \beta }} +
  Y^{{\theta}^2}\underscore9\,{r_j}\,{r_k}\,
   {{{{\gamma }^{ij}}}_{\alpha \beta }}\,
   {{{{\gamma }^{mik}}}_{\rho \sigma }} + 
  Y^{{\theta}^2}\underscore{10}\,{r_j}\,{r_k}\,
   {{{{\gamma }^{ij}}}_{\rho \sigma }}\,
   {{{{\gamma }^{mik}}}_{\alpha \beta }} \brkeq  + 
  Y^{{\theta}^2}\underscore{11}\,{r_l}\,{r_m}\,
   {{{{\gamma }^{ijk}}}_{\rho \sigma }}\,
   {{{{\gamma }^{ijkl}}}_{\alpha \beta }} +
  Y^{{\theta}^2}\underscore{12}\,{r_k}\,{r_l}\,
   {{{{\gamma }^{ijk}}}_{\rho \sigma }}\,
   {{{{\gamma }^{mijl}}}_{\alpha \beta }}  ,
\end{align}
where $Y^{{\theta}^2}_i $ $(i = 1 \sim 12)$ 
are  functions of $r(\tau)$ only. 
SUSY transformation laws then become
\begin{align}
\Dept r^m  &= {Y_1}\,(r \cdot v)\,(\epsilon {{\gamma }^{m}}{\dot{\theta}}
    ) - {Y_2}\,{r_i}\,{v_j}\,
   (\epsilon {{\gamma }^{mij}}{\dot{\theta}})
+ Y_{m \alpha\beta\rho\sigma}^{\theta^2} \epsilon_\beta \dot \theta_\alpha
\theta_\rho \theta_\sigma  , 
\\
\Dept \theta^\alpha &=
 {Y_1}\,{a_i}\,(r \cdot v)\,
   {{({{\gamma }^{i}}\epsilon )}_{\alpha }} + 
  {Y_2}\,{r_k}\,{v_j}\,{a_i}\,
   {{({{\gamma }^{ijk}}\epsilon )}_{\alpha }}
+ Y_{m \alpha\beta\rho\sigma}^{\theta^2} \epsilon_\beta 
\theta_\rho \theta_\sigma  a_m .  
\end{align}
Now we note that there exists a field redefinition of the form 
\begin{align}
\tilde r^m = r^m, \ \ \ \ 
\tilde \theta_\alpha = \theta_\alpha 
+ i Y_1 (r \cdot v) \dot \theta_\alpha .
\end{align}
which preserves the form of the effective action. 
It, however, changes the form of the  transformation laws as
\begin{align}
 \delta_\epsilon^{(2)} \tilde r^m  &=   \delta_\epsilon^{(2)} r^m -
{Y_1}\,(r \cdot v)\,(\epsilon {{\gamma }^{m}}{\dot{\theta}}) ,\\
 \delta_\epsilon^{(2)} \tilde \theta_\alpha  &=   
\delta_\epsilon^{(2)} \theta_\alpha -
 {Y_1}\,{a_i}\,(r \cdot v)\,{{({{\gamma }^{i}}\epsilon )}_{\alpha }} .
\end{align}
Thus, by using this field redefinition, we can always set $Y_1 = 0$.
With this choice, the additional contribution to 
the closure relation for $\theta_\alpha$ becomes 
\begin{align}
\left( \left[ \depz, \Dlamt\right] -(\ep \leftrightarrow \lam)
\right)\theta_\alpha & = \imag \,{Y_2}\,{r_j}\,{v_k}\,
   (\epsilon {{\gamma }^{i}}{\ddot{\theta}})\,
   {{(\lambda {{\gamma }^{ijk}})}_{\alpha }} + 
  \imag \,{Y_2}\,{r_j}\,{v_k}\,
   (\epsilon {{\gamma }^{ijk}}{\ddot{\theta}})\,
   {{(\lambda {{\gamma }^{i}})}_{\alpha }} \brkeq 
- 
  \imag \,{Y_2}\,{r_j}\,{v_k}\,
   (\lambda {{\gamma }^{i}}{\ddot{\theta}})\,
   {{(\epsilon {{\gamma }^{ijk}})}_{\alpha }} - 
  \imag \,{Y_2}\,{r_j}\,{v_k}\,
   (\lambda {{\gamma }^{ijk}}{\ddot{\theta}})\,
   {{(\epsilon {{\gamma }^{i}})}_{\alpha }}  \nn \\
&  - i (\epsilon \gamma^m \ddot \theta)
Y_{m \alpha\beta\rho\sigma}^{\theta^2} \lambda_\beta 
\theta_\rho \theta_\sigma 
+ i (\lambda \gamma^m \ddot \theta)
Y_{m \alpha\beta\rho\sigma}^{\theta^2} \epsilon_\beta 
\theta_\rho \theta_\sigma \nn \\
& - i Y_{m \alpha\beta\rho\sigma}^{\theta^2} \lambda_\beta 
\ddot \theta_\alpha 
\theta_\rho \theta_\sigma 
(\gamma^m \epsilon)_\alpha
+ i Y_{m \alpha\beta\rho\sigma}^{\theta^2} \epsilon_\beta 
\ddot \theta_\alpha 
\theta_\rho \theta_\sigma 
(\gamma^m \lambda)_\alpha \nn \\
& + \text{terms with $a_m$ and $\dot \theta_\alpha$.} 
\end{align}
As before, the terms with $a_m$ and $\dot \theta_\alpha$ 
only produce changes in the off-shell
coefficient $A_{\alpha\beta\gamma\delta}$, $B_{\alpha\beta\gamma n}$. 
On the other hand,
 the first 8 terms,  which do not vanish by any use of the Fierz
identities,  contain $\dot a^m$ and cannot be absorbed by  the
off-shell coefficients. Thus, the solution (ii) does not lead to proper SUSY
       transformation laws.
\item \quad By using the SUSY transformation laws
(\ref{case3r}) and  (\ref{case3th}), we can easily compute 
 the extra term produced in the closure
relation on $r^m$ to be $-8\,k_1\,(\epsilon \lambda )\,v^2\,{v_m}$.  
Proper closure relation cannot contain
such a term and hence the case (iii) is also excluded. 
\item \quad Finally we come to the case (iv).  
The additional terms produced in the closure relation on $r^m$ take the 
 form 
\begin{align}
 \and 8\,i\,{k_2}\,v^2\,(\epsilon {{\gamma }^i}\theta )\,
     (\lambda {{\gamma }^j}\theta )\,
     (\psi {{\gamma }^{ij}}\theta )-
    4\,i\,{k_2}\,{v_i}\,{v_j}\,
     (\epsilon {{\gamma }^{ikl}}\theta )\,
     (\lambda {{\gamma }^k}\theta )\,
     (\psi {{\gamma }^{jl}}\theta ) \brkeq 
   +4\,i\,{k_2}\,{v_i}\,{v_j}\,
     (\epsilon {{\gamma }^k}\theta )\,
     (\lambda {{\gamma }^{ikl}}\theta )\,
     (\psi {{\gamma }^{jl}}\theta )+
    4\,i\,{k_2}\,v^2\,(\epsilon {{\gamma }^i}\theta )\,
     (\lambda {{\gamma }^j}\psi )\,
     (\theta {{\gamma }^{ij}}\theta ) \brkeq 
   +4\,i\,{k_2}\,v^2\,(\epsilon {{\gamma }^i}\psi )\,
     (\lambda {{\gamma }^j}\theta )\,
     (\theta {{\gamma }^{ij}}\theta )+
    8\,i\,{k_2}\,{v_i}\,{v_j}\,(\epsilon \lambda )\,
     (\psi {{\gamma }^{ik}}\theta )\,
     (\theta {{\gamma }^{jk}}\theta ) \brkeq 
   -2\,i\,{k_2}\,{v_i}\,{v_j}\,
     (\epsilon {{\gamma }^{ikl}}\theta )\,
     (\lambda {{\gamma }^k}\psi )\,
     (\theta {{\gamma }^{jl}}\theta )-
    2\,i\,{k_2}\,{v_i}\,{v_j}\,
     (\epsilon {{\gamma }^{ikl}}\psi )\,
     (\lambda {{\gamma }^k}\theta )\,
     (\theta {{\gamma }^{jl}}\theta ) \brkeq 
   +2\,i\,{k_2}\,{v_i}\,{v_j}\,
     (\epsilon {{\gamma }^k}\theta )\,
     (\lambda {{\gamma }^{ikl}}\psi )\,
     (\theta {{\gamma }^{jl}}\theta )+
    2\,i\,{k_2}\,{v_i}\,{v_j}\,
     (\epsilon {{\gamma }^k}\psi )\,
     (\lambda {{\gamma }^{ikl}}\theta )\,
     (\theta {{\gamma }^{jl}}\theta ) . 
\end{align}
They are of unallowed form and it is not difficult to show
 that they do not vanish by any of 
the Fierz identities. Thus we must discard this final possibility. 
\end{enumerate}

This demonstrates that there are no acceptable null transformations 
 and hence the D-type basis  is independent. This in turn 
 is responsible for the uniqueness of the 
 the SUSY transformation laws  determined by the SUSY Ward identities.

\newpage
\setcounter{equation}{0}
\renewcommand{\theequation}{B.\arabic{equation}}
\section*{Appendix B: \ \  SUSY transformation laws}

In Sec.~4.3, we recorded  $\calO(\theta^0)$ and  $\calO(\theta^2)$
parts of the SUSY transformation laws. 
In this appendix,  we display the remaining $\calO(\theta^4)$ and 
 $\calO(\theta^6)$ parts.

\bigskip
\noindent
{\bf\boldmath  $\calO(\theta^4)$ part:}
\begin{align}
\Omega_{m \beta}^{\theta^5} \epsilon_\beta 
= \and \frac{7\,i\,c\,(\epsilon {{\gamma }^i}\theta )\,
       (\theta {{\gamma }^{ij}}\theta )\,
       (\theta {{\gamma }^{jm}}\theta )}{16\,r^9} \brkeq 
+ \frac{91\,i\,c\,{r_i}\,{r_m}\,
       (\epsilon {{\gamma }^j}\theta )\,
       (\theta {{\gamma }^{ik}}\theta )\,
       (\theta {{\gamma }^{jk}}\theta )}{80\,r^{11}}-
    \frac{49\,i\,c\,{r_i}\,{r_j}\,
       (\epsilon {{\gamma }^m}\theta )\,
       (\theta {{\gamma }^{ik}}\theta )\,
       (\theta {{\gamma }^{jk}}\theta )}{80\,r^{11}} \brkeq
   +\frac{119\,i\,c\,{r_i}\,{r_j}\,
       (\epsilon {{\gamma }^k}\theta )\,
       (\theta {{\gamma }^{ik}}\theta )\,
       (\theta {{\gamma }^{jm}}\theta )}{20\,r^{11}}-
    \frac{2793\,i\,c\,{r_i}\,{r_j}\,
       (\epsilon {{\gamma }^{jkm}}\theta )\,
       (\theta {{\gamma }^{il}}\theta )\,
       (\theta {{\gamma }^{kl}}\theta )}{640\,r^{11}}
    \brkeq +\frac{2513\,i\,c\,{r_i}\,{r_j}\,
       (\epsilon {{\gamma }^{jkl}}\theta )\,
       (\theta {{\gamma }^{il}}\theta )\,
       (\theta {{\gamma }^{km}}\theta )}{640\,r^{11}}+
    \frac{49\,i\,c\,{r_i}\,{r_j}\,
       (\epsilon {{\gamma }^i}\theta )\,
       (\theta {{\gamma }^{jk}}\theta )\,
       (\theta {{\gamma }^{km}}\theta )}{16\,r^{11}} \brkeq
    +\frac{2513\,i\,c\,{r_i}\,{r_j}\,
       (\epsilon {{\gamma }^{km}}\theta )\,
       (\theta {{\gamma }^{il}}\theta )\,
       (\theta {{\gamma }^{jkl}}\theta )}{640\,r^{11}}+
    \frac{21\,i\,c\,{r_i}\,{r_j}\,
       (\epsilon {{\gamma }^{ik}}\theta )\,
       (\theta {{\gamma }^{lm}}\theta )\,
       (\theta {{\gamma }^{jkl}}\theta )}{8\,r^{11}} \brkeq
    -\frac{119\,i\,c\,{r_i}\,{r_j}\,(\epsilon \theta )\,
       (\theta {{\gamma }^{ik}}\theta )\,
       (\theta {{\gamma }^{jkm}}\theta )}{80\,r^{11}}-
    \frac{2289\,i\,c\,{r_i}\,{r_j}\,
       (\epsilon {{\gamma }^{kl}}\theta )\,
       (\theta {{\gamma }^{il}}\theta )\,
       (\theta {{\gamma }^{jkm}}\theta )}{640\,r^{11}}
    \brkeq -\frac{189\,i\,c\,{r_i}\,{r_j}\,
       (\epsilon {{\gamma }^{ik}}\theta )\,
       (\theta {{\gamma }^{kl}}\theta )\,
       (\theta {{\gamma }^{jlm}}\theta )}{80\,r^{11}}+
    \frac{4641\,i\,c\,{r_i}\,{r_j}\,
       (\epsilon {{\gamma }^{ik}}\theta )\,
       (\theta {{\gamma }^{jl}}\theta )\,
       (\theta {{\gamma }^{klm}}\theta )}{640\,r^{11}} . 
\label{strp0th5}
\end{align}

\begin{align}
T_{\alpha\beta}^{\p \theta^4} \epsilon_\beta 
= \and -\frac{119\,i\,c\,{r_i}\,{r_j}\,{v_k}\,
       (\theta {{\gamma }^{il}}\theta )\,
       (\theta {{\gamma }^{jkl}}\theta )\,
       {{\epsilon }_{\alpha }}}{80\,r^{11}}+
    \frac{49\,i\,c\,{r_i}\,{r_j}\,{v_k}\,
       (\theta {{\gamma }^{jl}}\theta )\,
       (\theta {{\gamma }^{kl}}\theta )\,
       {{(\epsilon {{\gamma }^{i}})}_{\alpha }}}{16\,
       r^{11}} \brkeq -\frac{21\,i\,c\,{v_i}\,
       (\theta {{\gamma }^{ik}}\theta )\,
       (\theta {{\gamma }^{jk}}\theta )\,
       {{(\epsilon {{\gamma }^{j}})}_{\alpha }}}{16\,r^9}-
    \frac{91\,i\,c\,{r_i}\,(r \cdot v)\,
       (\theta {{\gamma }^{ik}}\theta )\,
       (\theta {{\gamma }^{jk}}\theta )\,
       {{(\epsilon {{\gamma }^{j}})}_{\alpha }}}{80\,
       r^{11}} \brkeq +\frac{49\,i\,c\,{r_i}\,{r_j}\,
       {v_k}\,(\theta {{\gamma }^{il}}\theta )\,
       (\theta {{\gamma }^{jl}}\theta )\,
       {{(\epsilon {{\gamma }^{k}})}_{\alpha }}}{80\,
       r^{11}}+\frac{49\,i\,c\,{r_i}\,{r_j}\,{v_k}\,
       (\theta {{\gamma }^{il}}\theta )\,
       (\theta {{\gamma }^{jk}}\theta )\,
       {{(\epsilon {{\gamma }^{l}})}_{\alpha }}}{5\,r^{11}}
    \brkeq -\frac{189\,i\,c\,{r_i}\,{r_j}\,{v_k}\,
       (\theta {{\gamma }^{lm}}\theta )\,
       (\theta {{\gamma }^{jkm}}\theta )\,
       {{(\epsilon {{\gamma }^{il}})}_{\alpha }}}{80\,
       r^{11}}-\frac{21\,i\,c\,{r_i}\,{r_j}\,{v_k}\,
       (\theta {{\gamma }^{kl}}\theta )\,
       (\theta {{\gamma }^{jlm}}\theta )\,
       {{(\epsilon {{\gamma }^{im}})}_{\alpha }}}{8\,
       r^{11}} \brkeq +\frac{4641\,i\,c\,{r_i}\,{r_j}\,
       {v_k}\,(\theta {{\gamma }^{jl}}\theta )\,
       (\theta {{\gamma }^{klm}}\theta )\,
       {{(\epsilon {{\gamma }^{im}})}_{\alpha }}}{640\,
       r^{11}}-\frac{2513\,i\,c\,{r_i}\,{r_j}\,{v_k}\,
       (\theta {{\gamma }^{il}}\theta )\,
       (\theta {{\gamma }^{jlm}}\theta )\,
       {{(\epsilon {{\gamma }^{km}})}_{\alpha }}}{640\,
       r^{11}} \brkeq +\frac{2289\,i\,c\,{r_i}\,{r_j}\,
       {v_k}\,(\theta {{\gamma }^{il}}\theta )\,
       (\theta {{\gamma }^{jkm}}\theta )\,
       {{(\epsilon {{\gamma }^{lm}})}_{\alpha }}}{640\,
       r^{11}}-\frac{2793\,i\,c\,{r_i}\,{r_j}\,{v_k}\,
       (\theta {{\gamma }^{im}}\theta )\,
       (\theta {{\gamma }^{lm}}\theta )\,
       {{(\epsilon {{\gamma }^{jkl}})}_{\alpha }}}{640\,
       r^{11}} \brkeq 
+\frac{2513\,i\,c\,{r_i}\,{r_j}\,
       {v_k}\,(\theta {{\gamma }^{im}}\theta )\,
       (\theta {{\gamma }^{kl}}\theta )\,
       {{(\epsilon {{\gamma }^{jlm}})}_{\alpha }}}{640\,
       r^{11}}-\frac{1841\,i\,c\,{r_i}\,{r_j}\,{v_k}\,
       (\theta {{\gamma }^{il}}\theta )\,
       (\theta {{\gamma }^{jm}}\theta )\,
       {{(\epsilon {{\gamma }^{klm}})}_{\alpha }}}{640\,
       r^{11}} \brkeq +\frac{7\,i\,c\,{v_i}\,
       (\epsilon {{\gamma }^j}\theta )\,
       (\theta {{\gamma }^{jk}}\theta )\,
       {{(\theta {{\gamma }^{ik}})}_{\alpha }}}{8\,r^9}-
    \frac{91\,i\,c\,{r_i}\,(r \cdot v)\,
       (\epsilon {{\gamma }^j}\theta )\,
       (\theta {{\gamma }^{jk}}\theta )\,
       {{(\theta {{\gamma }^{ik}})}_{\alpha }}}{40\,r^{11}}
    \brkeq 
-\frac{119\,i\,c\,{r_i}\,{r_j}\,{v_k}\,
       (\epsilon {{\gamma }^l}\theta )\,
       (\theta {{\gamma }^{jk}}\theta )\,
       {{(\theta {{\gamma }^{il}})}_{\alpha }}}{10\,r^{11}}
     -\frac{1841\,i\,c\,{r_i}\,{r_j}\,{v_k}\,
       (\epsilon {{\gamma }^{klm}}\theta )\,
       (\theta {{\gamma }^{jm}}\theta )\,
       {{(\theta {{\gamma }^{il}})}_{\alpha }}}{320\,
       r^{11}} \brkeq 
-\frac{119\,i\,c\,{r_i}\,{r_j}\,
       {v_k}\,(\epsilon \theta )\,
       (\theta {{\gamma }^{jkl}}\theta )\,
       {{(\theta {{\gamma }^{il}})}_{\alpha }}}{40\,r^{11}}
     +\frac{2513\,i\,c\,{r_i}\,{r_j}\,{v_k}\,
       (\epsilon {{\gamma }^{jlm}}\theta )\,
       (\theta {{\gamma }^{kl}}\theta )\,
       {{(\theta {{\gamma }^{im}})}_{\alpha }}}{320\,
       r^{11}} \nn \end{align}
\begin{align}
\brkeq -\frac{2793\,i\,c\,{r_i}\,{r_j}\,
       {v_k}\,(\epsilon {{\gamma }^{jkl}}\theta )\,
       (\theta {{\gamma }^{lm}}\theta )\,
       {{(\theta {{\gamma }^{im}})}_{\alpha }}}{320\,
       r^{11}}-\frac{2289\,i\,c\,{r_i}\,{r_j}\,{v_k}\,
       (\epsilon {{\gamma }^{lm}}\theta )\,
       (\theta {{\gamma }^{jkl}}\theta )\,
       {{(\theta {{\gamma }^{im}})}_{\alpha }}}{320\,
       r^{11}} \brkeq +\frac{2513\,i\,c\,{r_i}\,{r_j}\,
       {v_k}\,(\epsilon {{\gamma }^{kl}}\theta )\,
       (\theta {{\gamma }^{jlm}}\theta )\,
       {{(\theta {{\gamma }^{im}})}_{\alpha }}}{320\,
       r^{11}}+\frac{7\,i\,c\,{v_i}\,
       (\epsilon {{\gamma }^j}\theta )\,
       (\theta {{\gamma }^{ik}}\theta )\,
       {{(\theta {{\gamma }^{jk}})}_{\alpha }}}{8\,r^9}
    \brkeq -\frac{91\,i\,c\,{r_i}\,(r \cdot v)\,
       (\epsilon {{\gamma }^j}\theta )\,
       (\theta {{\gamma }^{ik}}\theta )\,
       {{(\theta {{\gamma }^{jk}})}_{\alpha }}}{40\,r^{11}}
     -\frac{119\,i\,c\,{r_i}\,{r_j}\,{v_k}\,
       (\epsilon {{\gamma }^l}\theta )\,
       (\theta {{\gamma }^{il}}\theta )\,
       {{(\theta {{\gamma }^{jk}})}_{\alpha }}}{10\,r^{11}}
    \brkeq +\frac{49\,i\,c\,{r_i}\,{r_j}\,{v_k}\,
       (\epsilon {{\gamma }^k}\theta )\,
       (\theta {{\gamma }^{il}}\theta )\,
       {{(\theta {{\gamma }^{jl}})}_{\alpha }}}{20\,r^{11}}
     +\frac{49\,i\,c\,{r_i}\,{r_j}\,{v_k}\,
       (\epsilon {{\gamma }^i}\theta )\,
       (\theta {{\gamma }^{kl}}\theta )\,
       {{(\theta {{\gamma }^{jl}})}_{\alpha }}}{8\,r^{11}}
    \brkeq -\frac{1841\,i\,c\,{r_i}\,{r_j}\,{v_k}\,
       (\epsilon {{\gamma }^{klm}}\theta )\,
       (\theta {{\gamma }^{il}}\theta )\,
       {{(\theta {{\gamma }^{jm}})}_{\alpha }}}{320\,
       r^{11}}-\frac{4641\,i\,c\,{r_i}\,{r_j}\,{v_k}\,
       (\epsilon {{\gamma }^{il}}\theta )\,
       (\theta {{\gamma }^{klm}}\theta )\,
       {{(\theta {{\gamma }^{jm}})}_{\alpha }}}{320\,
       r^{11}} \brkeq +\frac{2513\,i\,c\,{r_i}\,{r_j}\,
       {v_k}\,(\epsilon {{\gamma }^{jlm}}\theta )\,
       (\theta {{\gamma }^{im}}\theta )\,
       {{(\theta {{\gamma }^{kl}})}_{\alpha }}}{320\,
       r^{11}}+\frac{49\,i\,c\,{r_i}\,{r_j}\,{v_k}\,
       (\epsilon {{\gamma }^i}\theta )\,
       (\theta {{\gamma }^{jl}}\theta )\,
       {{(\theta {{\gamma }^{kl}})}_{\alpha }}}{8\,r^{11}}
    \brkeq +\frac{21\,i\,c\,{r_i}\,{r_j}\,{v_k}\,
       (\epsilon {{\gamma }^{il}}\theta )\,
       (\theta {{\gamma }^{jlm}}\theta )\,
       {{(\theta {{\gamma }^{km}})}_{\alpha }}}{4\,r^{11}}-
    \frac{2793\,i\,c\,{r_i}\,{r_j}\,{v_k}\,
       (\epsilon {{\gamma }^{jkl}}\theta )\,
       (\theta {{\gamma }^{im}}\theta )\,
       {{(\theta {{\gamma }^{lm}})}_{\alpha }}}{320\,
       r^{11}} \brkeq -\frac{189\,i\,c\,{r_i}\,{r_j}\,
       {v_k}\,(\epsilon {{\gamma }^{il}}\theta )\,
       (\theta {{\gamma }^{jkm}}\theta )\,
       {{(\theta {{\gamma }^{lm}})}_{\alpha }}}{40\,r^{11}}
     -\frac{119\,i\,c\,{r_i}\,{r_j}\,{v_k}\,
       (\epsilon \theta )\,
       (\theta {{\gamma }^{il}}\theta )\,
       {{(\theta {{\gamma }^{jkl}})}_{\alpha }}}{40\,
       r^{11}} \brkeq -\frac{2289\,i\,c\,{r_i}\,{r_j}\,
       {v_k}\,(\epsilon {{\gamma }^{lm}}\theta )\,
       (\theta {{\gamma }^{im}}\theta )\,
       {{(\theta {{\gamma }^{jkl}})}_{\alpha }}}{320\,
       r^{11}}-\frac{189\,i\,c\,{r_i}\,{r_j}\,{v_k}\,
       (\epsilon {{\gamma }^{il}}\theta )\,
       (\theta {{\gamma }^{lm}}\theta )\,
       {{(\theta {{\gamma }^{jkm}})}_{\alpha }}}{40\,
       r^{11}} \brkeq +\frac{2513\,i\,c\,{r_i}\,{r_j}\,
       {v_k}\,(\epsilon {{\gamma }^{kl}}\theta )\,
       (\theta {{\gamma }^{im}}\theta )\,
       {{(\theta {{\gamma }^{jlm}})}_{\alpha }}}{320\,
       r^{11}}+\frac{21\,i\,c\,{r_i}\,{r_j}\,{v_k}\,
       (\epsilon {{\gamma }^{il}}\theta )\,
       (\theta {{\gamma }^{km}}\theta )\,
       {{(\theta {{\gamma }^{jlm}})}_{\alpha }}}{4\,r^{11}}
    \brkeq -\frac{4641\,i\,c\,{r_i}\,{r_j}\,{v_k}\,
       (\epsilon {{\gamma }^{il}}\theta )\,
       (\theta {{\gamma }^{jm}}\theta )\,
       {{(\theta {{\gamma }^{klm}})}_{\alpha }}}{320\,
       r^{11}} .
\label{stthp1th4}
\end{align}

\bigskip
\noindent
{\bf\boldmath  $\calO(\theta^6)$ part:}
\begin{align}
T_{\alpha\beta}^{\theta^6} \epsilon_\beta =
  \and \frac{27\,i\,c\,{r_i}\,
       (\theta {{\gamma }^{il}}\theta )\,
       (\theta {{\gamma }^{jk}}\theta )\,
       (\theta {{\gamma }^{kl}}\theta )\,
       {{(\epsilon {{\gamma }^{j}})}_{\alpha }}}{40\,
       r^{11}}+\frac{143\,i\,c\,{r_i}\,{r_j}\,{r_k}\,
       (\theta {{\gamma }^{il}}\theta )\,
       (\theta {{\gamma }^{jm}}\theta )\,
       (\theta {{\gamma }^{km}}\theta )\,
       {{(\epsilon {{\gamma }^{l}})}_{\alpha }}}{40\,
       r^{13}} \brkeq +\frac{i\,c\,{r_i}\,
       (\theta {{\gamma }^{jk}}\theta )\,
       (\theta {{\gamma }^{kl}}\theta )\,
       (\theta {{\gamma }^{lm}}\theta )\,
       {{(\epsilon {{\gamma }^{ijm}})}_{\alpha }}}{80\,
       r^{11}}-\frac{11\,i\,c\,{r_i}\,{r_j}\,{r_k}\,
       (\theta {{\gamma }^{il}}\theta )\,
       (\theta {{\gamma }^{jn}}\theta )\,
       (\theta {{\gamma }^{mn}}\theta )\,
       {{(\epsilon {{\gamma }^{klm}})}_{\alpha }}}{80\,
       r^{13}} \brkeq -\frac{2\,i\,c\,{r_i}\,
       (\epsilon {{\gamma }^j}\theta )\,
       (\theta {{\gamma }^{jk}}\theta )\,
       (\theta {{\gamma }^{kl}}\theta )\,
       {{(\theta {{\gamma }^{il}})}_{\alpha }}}{5\,r^{11}}-
    \frac{99\,i\,c\,{r_i}\,{r_j}\,{r_k}\,
       (\epsilon {{\gamma }^l}\theta )\,
       (\theta {{\gamma }^{jm}}\theta )\,
       (\theta {{\gamma }^{km}}\theta )\,
       {{(\theta {{\gamma }^{il}})}_{\alpha }}}{40\,r^{13}}
    \brkeq -\frac{11\,i\,c\,{r_i}\,{r_j}\,{r_k}\,
       (\epsilon {{\gamma }^{klm}}\theta )\,
       (\theta {{\gamma }^{jn}}\theta )\,
       (\theta {{\gamma }^{mn}}\theta )\,
       {{(\theta {{\gamma }^{il}})}_{\alpha }}}{40\,r^{13}}
     -\frac{2\,i\,c\,{r_i}\,
       (\epsilon {{\gamma }^j}\theta )\,
       (\theta {{\gamma }^{il}}\theta )\,
       (\theta {{\gamma }^{kl}}\theta )\,
       {{(\theta {{\gamma }^{jk}})}_{\alpha }}}{5\,r^{11}}
    \brkeq -\frac{i\,c\,{r_i}\,
       (\epsilon {{\gamma }^{ijk}}\theta )\,
       (\theta {{\gamma }^{km}}\theta )\,
       (\theta {{\gamma }^{lm}}\theta )\,
       {{(\theta {{\gamma }^{jl}})}_{\alpha }}}{40\,r^{11}}
     -\frac{11\,i\,c\,{r_i}\,{r_j}\,{r_k}\,
       (\epsilon {{\gamma }^{klm}}\theta )\,
       (\theta {{\gamma }^{il}}\theta )\,
       (\theta {{\gamma }^{mn}}\theta )\,
       {{(\theta {{\gamma }^{jn}})}_{\alpha }}}{40\,r^{13}}
    \brkeq -\frac{2\,i\,c\,{r_i}\,
       (\epsilon {{\gamma }^j}\theta )\,
       (\theta {{\gamma }^{il}}\theta )\,
       (\theta {{\gamma }^{jk}}\theta )\,
       {{(\theta {{\gamma }^{kl}})}_{\alpha }}}{5\,r^{11}}-
    \frac{99\,i\,c\,{r_i}\,{r_j}\,{r_k}\,
       (\epsilon {{\gamma }^l}\theta )\,
       (\theta {{\gamma }^{il}}\theta )\,
       (\theta {{\gamma }^{jm}}\theta )\,
       {{(\theta {{\gamma }^{km}})}_{\alpha }}}{20\,r^{13}}
    \brkeq -\frac{i\,c\,{r_i}\,
       (\epsilon {{\gamma }^{ijk}}\theta )\,
       (\theta {{\gamma }^{jl}}\theta )\,
       (\theta {{\gamma }^{lm}}\theta )\,
       {{(\theta {{\gamma }^{km}})}_{\alpha }}}{40\,r^{11}}
     -\frac{i\,c\,{r_i}\,
       (\epsilon {{\gamma }^{ijk}}\theta )\,
       (\theta {{\gamma }^{jl}}\theta )\,
       (\theta {{\gamma }^{km}}\theta )\,
       {{(\theta {{\gamma }^{lm}})}_{\alpha }}}{40\,r^{11}}
    \brkeq -\frac{11\,i\,c\,{r_i}\,{r_j}\,{r_k}\,
       (\epsilon {{\gamma }^{klm}}\theta )\,
       (\theta {{\gamma }^{il}}\theta )\,
       (\theta {{\gamma }^{jn}}\theta )\,
       {{(\theta {{\gamma }^{mn}})}_{\alpha }}}{40\,r^{13}} .
\label{stthp0th6}
\end{align}


\end{document}